\documentclass[aps,prd]{revtex4}
\usepackage[latin1]{inputenc}
\usepackage{amsmath}
\usepackage{amsfonts}
\usepackage{amssymb}
\usepackage{amsthm}

\ifx\pdftexversion\undefined
 \usepackage[dvips]{graphics}
\fi
\usepackage[pdftex]{graphicx}

\newcommand{\diff}{\mathrm{d}}
\newtheorem{thm}{Theorem}[section]

\theoremstyle{remark}

\theoremstyle{definition}
\newtheorem{defn}[thm]{Definition}

\begin{document}

\title{The Local Effects of Cosmological Variations in Physical 'Constants'
and Scalar Fields \\
I. Spherically Symmetric Spacetimes}
\author{Douglas J. Shaw}
\affiliation{DAMTP, Centre for Mathematical Sciences, University of Cambridge,
Wilberforce Road, Cambridge CB3 0WA, UK}
\author{John D. Barrow}
\affiliation{DAMTP, Centre for Mathematical Sciences, University of Cambridge,
Wilberforce Road, Cambridge CB3 0WA, UK}
\date{4th December 2005}

\begin{abstract}
We apply the method of matched asymptotic expansions to analyse whether
cosmological variations in physical `constants' and scalar fields are
detectable, locally, on the surface of local gravitationally bound systems
such as planets and stars, or inside virialised systems like galaxies and
clusters. We assume spherical symmetry and derive a sufficient condition for
the local time variation of the scalar fields that drive varying constants
to track the cosmological one. We calculate a number of specific
examples in detail by matching the Schwarzschild spacetime to spherically
symmetric inhomogeneous Tolman-Bondi metrics in an intermediate region by
rigorously construction matched asymptotic expansions on cosmological and
local astronomical scales which overlap in an intermediate domain. We
conclude that, independent of the details of the scalar-field theory
describing the varying `constant', the condition for cosmological variations
to be measured locally is almost always satisfied in physically realistic
situations. The proof of this statement provides a rigorous justification
for using terrestrial experiments and solar system observations to
constrain or detect any cosmological time variations in the traditional
`constants' of Nature.

\end{abstract}

\maketitle

\section{\protect\bigskip Introduction}

In recent years there has been growing interest in the observational and
theoretical consequences of time variations in the values of the traditional
constants of Nature, notably of the fine structure constant, $\alpha $, \cite%
{webb, chand, sdss, qu, lev, lev2, rocha, darl, oh, drink}, the
electron-proton mass ratio, $\mu =m_{e}/m_{pr}$, \cite{ubachs, petit, tz},
and Newton's `constant' of gravitation, $G$, \cite{bert}. In all cases the
experimental evidence that can be brought to bear on the problem is a
combination of local (laboratory, terrestrial, and solar system) and global
(astronomical and cosmological) observations \cite{uzan, olive, jdb,
jdbroysoc, posp}. The first theoretical challenge is to develop
self-consistent extensions of general relativity which incorporate varying
`constants' rigorously by including the gravitational effects of the
variations and ensuring that energy and momentum are totally conserved by
the variations that replace the former constants. This is achieved by
regarding the `constant' as a scalar field with particular couplings and
self-interaction. In the case of varying $G$, the Brans-Dicke theory \cite%
{bd, bkm} provides the paradigm for a scalar-tensor theory of this type.
Recently, the same philosophy has been applied to produce simple extensions
of general relativity which self-consistently describe the spacetime
variation of $\alpha $, \cite{bek, bsbm}, and $\mu $, \cite{bmmu}. It is
also possible to extend these studies to include simultaneous variations of
several constants, and to include the weak coupling by a generalisation of
the Weinberg-Salam theory to include spacetime varying coupling `constants'
\cite{bs, sb}. In these theories the analysis of the behaviour of their
solutions is simplified because we know that the allowed variations in
constants like $\alpha $ are constrained already to be `small' and will
not have any significant effect on the expansion dynamics of the universe in
recent times. The Brans-Dicke theory is different. Small variations in $G$
will always have direct consequences for the expansion dynamics of the
universe. Typically, a power-law time variation of $G\propto t^{-n}$ creates
a variation of the expansion scale factor that goes as $a(t)\propto
t^{(2-n)/3}$ in a dust-dominated Friedmann universe \cite{newtgrav}.

These theories are confronted with a variety of laboratory, geochemical, and
astronomical observations. In the case of variations of $\alpha $ we have
laboratory constraints on atomic lines, indirect bounds from the Oklo
natural reactor operation 1.8 billion years ago, radioactive decay products
in meteoritic data back to 4.6 billion years ago, and quasar spectra out to
redshifts $z\lesssim O(4)$, as the prime sources of observational evidence
against which to test theories which permit spacetime variations \cite{uzan,
olive, jdb, jdbroysoc, posp}. Generally, the data from all these diverse
physical scales are lumped together and used to test time variations of $%
\alpha $. Thus laboratory or solar system evidence is compared directly with
quasar data and used to constrain the allowed cosmological variations of $%
\alpha $. Similar tactics are used to constrain the allowed variations of
other constants, like $G$ or $\mu $.

This simultaneous use of terrestrial and astronomical bounds on constants
assumes implicitly the unproven requirement that any variation of a constant
on cosmological scales is `seen' locally inside virialised structures like
galaxies or solar systems, and has a measurable effect in laboratory
experiments on Earth. It is not obvious a priori that this need be the case:
we would not expect to test the expansion of the universe by measuring the
expansion of the Earth. The central question that this paper addresses is
the extent to which global variations of `constants' on cosmological scales
that take part in the Hubble expansion of the universe are seen locally on
the surface of gravitationally-bound structures, like planets, or inside
bound systems of stars like galaxies \cite{bmot}. Only if we can show that
cosmological variations have calculable local effects will it be legitimate
to use laboratory and solar system observations to constrain theories of
varying constants in the way that is habitually done, without proof, in the
literature. So far, detailed analyses of spatial variations of constants
have been made only for small variations, where the isotropy of the
microwave background places very strong limits on spatial variations because
of the associated Sachs-Wolfe effects created by the gravitational potential
perturbations that accompany spatial fluctuations in `constants' via their
associated scalar fields because of the coupling of the latter to matter
\cite{jbspace1, jbspace}.

We are concerned with the dynamics of spacetime scalar fields that are
weakly coupled to gravity and matter. We will not consider theories where
there are two or more scalar fields interacting amongst themselves, although
the method we use here could also be easily extended to that scenario.
Theories which introduce varying constants self-consistently into Einstein's
conception of a gravitation theory do so by associating the `constant' $\mathbb{C}$ to be varied with a new scalar field, so $\mathbb{C} \rightarrow \mathbb{C}(\varphi )$. The variations of this scalar field gravitate and contribute to
the curvature of spacetime like any other form of mass-energy. They must
also conserve energy and momentum and so their forms are constrained by a
covariant conservation equation for the scalar field. Typically, this
results in a wave equation of the form
\begin{equation}
\square \varphi =\lambda f(\varphi )L(\rho ,p),  \label{gen}
\end{equation}%
where $\varphi $ is a scalar field associated with the variation of some
`constant' $\mathbb{C}$ via a relation $\mathbb{C}=f(\varphi )$,$\lambda $ is a dimensionless measure of the strength of the space-time variation of $\mathbb{C}$, $f(\varphi )$ is a function determined by the definition of $\varphi ,$
and $L(\rho ,p)$ is some linear combination of the density, $\rho $, and
pressure, $p$, of the matter that is coupled to the field $\varphi $ and $%
f(\varphi )\simeq 1$ for small variations in $\varphi $ and $\mathbb{C}$. This form includes all the standard theories for varying constants, like $%
G$, $\alpha $, and $\mu $, of refs. \cite{bd, bek, bsbm, bmmu}.

We shall refer to our scalar field as the `dilaton', denote it by $\phi $,
and analyse the form of the standard equation (\ref{gen}) a little further
by assuming that $\phi (\vec{x},t)$ satisfies a conservation equation that
can be decomposed into the form
\begin{equation}
\square \phi =B_{,\phi }(\phi )\kappa T-V_{,\phi }\left( \phi \right)
\label{cons}
\end{equation}%
where $T$ is the trace of the energy momentum tensor, $T=T_{\mu }^{\mu }$
(with the contribution from any cosmological constant neglected). We absorb
any dilaton-to-cosmological constant coupling into the definition of $V(\phi
)$. The dilaton to matter coupling $B(\phi )$ and the self-interaction
potential, $V(\phi )$, are arbitrary functions of $\phi $ and $\kappa =8\pi G
$ and $c=\hslash =1$. This covers a wide range of theories which describe
the spacetime variation of `constants' of Nature, and includes
Einstein-frame Brans-Dicke (BD) and all other, single field, scalar-tensor
theories of gravity. In cosmologies that are composed of dust, cosmological
constant and radiation it will also contain the
Bekenstein-Sandvik-Barrow-Magueijo (BSBM) of varying $\alpha $, \cite{bsbm},
and other (single dilaton) theories which describe the variation of standard
model couplings, \cite{posp}. Note that one could, for example, generalise
the form of this conservation equation, (\ref{cons}), whilst maintaining
relativistic invariance, by adding a coupling to $\sqrt{T^{\alpha \beta
}T_{\alpha \beta }}$, or we could also break local Lorentz invariance by
adding extra couplings to the pressure, $P$, defined w.r.t. to some
preferred coordinate system. In this paper we will mostly be considering
spacetimes in which the pressure of the matter vanishes, $P=0$, and so all
of the potential extra couplings mentioned above will reduce to a form that
is included in the conservation equation (\ref{cons}) that we assume.

We will assume that the background cosmology is isotropic and homogeneous
and work with a Friedmann-Robertson-Walker (FRW) background metric:
\begin{equation}
\mathrm{d}s^{2}=\mathrm{d}t^{2}-a^{2}(t)\left( \frac{\mathrm{d}r^{2}}{%
1-kr^{2}}+r^{2}\{\mathrm{d}\theta ^{2}+\sin ^{2}\theta \mathrm{d}\phi
^{2}\}\right) ,  \label{frw}
\end{equation}%
where $k$ is the spatial curvature parameter. In the background universe the
dilaton field is also assumed to be homogeneous (so $\phi =\phi _{c}(t)$)
and therefore satisfies the ordinary differential equation (ODE):
\begin{equation}
\frac{1}{a^{3}}\left( a^{3}\dot{\phi}_{c}\right) ^{\cdot }=B_{,\phi }(\phi
_{c})\kappa \left( \epsilon _{c}-3P_{c}\right) -V_{,\phi }\left( \phi
_{c}\right) .  \label{dilcons}
\end{equation}

\ \ The dilaton conservation equation also reduces to a similar ODE if we
are only interested in its time-independent mode $\phi (r$) in a static
spherically symmetric background, like the Schwarzschild metric. In the
actual universe, however, spacetimes that look static in some locality must
still match smoothly on to the cosmological background spacetime on large
scales. Hence, if we are to model the evolution of the dilaton field
accurately in some inhomogeneous region, embedded in a homogeneous
background, we must demand that at large distances $\phi \rightarrow \phi
_{c} $ in some appropriate way. Even if spherical symmetry is assumed, the
conservation equation is generically a non-linear, second-order, partial
differential equation (PDE) and its solution is far from straightforward to
find, either exactly or approximately. Even numerical models are technically
difficult to set-up, see ref. \cite{harada}, and only allow us to consider
one particular choice of $B(\phi )$, $V(\phi ),$ and the spacetime
background at a time. However, as we have mentioned above, if we are to
bring all of our experimental evidence to bear on these models, we need to
know how to interpret local observations in the light of our cosmological
ones. Perhaps the most important piece of knowledge we would like to have is
the correlation between the local and global (or cosmological) time
variation of the dilaton. In particular, we would like to find the criteria
under which it is true that
\begin{equation}
\dot{\phi}(\mathbf{x},t)\approx \dot{\phi}_{c}(t),  \label{wettcond}
\end{equation}%
that is, when does the local time-variation of $\phi $ track the
cosmological one?

In this paper we will try to answer this question by applying asymptotic
methods commonly used in fluid dynamics in order to construct asymptotic
approximations to the behaviour of $\phi $ close to a spherical static mass,
that match to the cosmological solution, $\phi _{c}$, at large distances.
From this analysis, we can derive a sufficient condition for eq. (\ref%
{wettcond}) to hold. We limit ourselves in this paper to considering
spacetime backgrounds that are a spherical symmetric. In a subsequent work
we will generalise our results to deal with more general non-spherical
backgrounds, \cite{shawbarrow2}.

Throughout our analysis we will refer to the spherical, static mass as a
`star', however it could be taken to be a planet (e.g. the Earth), a
black-hole, or even a galaxy or cluster of galaxies. We are mostly
interested in the (realistic) case where the surface of our `star' lies
far-outside its own Schwarzschild-radius. By applying our results to
the black-hole case, however, we will comment on the problem of `gravitational
memory'; that is, is the cosmological background value of $\phi $ on the
horizon at the time when a black hole forms, frozen-in, or `remembered',
when the black-hole forms, or does it continue to track the background
cosmological evolution?

This paper is organised as follows: in section II we review some of the
previous studies into the problem of local vs. global dilaton evolution and
note where our work extends and improves these studies. In section III we
will introduce the method of matched asymptotic expansions which we will
use to carry out our study, and provide some simple examples of its
application. In section IV we define our geometrical set-up of a star in a
background cosmology, and detail our particular choices of possible
spacetime backgrounds. Then, in section V, we construct overlapping
asymptotic expansions to study the constraints required if the local and
global evolution is to match together. In section VI we derive the
conditions that must be satisfied for our method to be valid. Taking these
into account, we interpret and generalise our results in section VII. In
this section we shall also make a conjecture about a general condition that
is sufficient for eq. (\ref{wettcond}) to hold, which will apply to more
general spacetime backgrounds that those explicitly considered in this
paper. Finally, in section VIII we use our sufficient condition to show that
we \emph{do} expect eq. (\ref{wettcond}) to hold here on Earth.

\section{Past Work}

Several authors have, in the past, claimed to have shown that condition (\ref%
{wettcond}) holds close to the surface at, $r=R_{s}$, of a
spherically-symmetric, massive body embedded, in some particular way, into
an expanding universe. The results derived are usually only valid for a
particular choice of $B(\phi )$ and $V(\phi )$ and under some, usually
restrictive, assumptions about the background distribution of matter. In
this section we will review the pioneering analyses carried out by Wetterich
in \cite{wetterich:2002} and Jacobson in \cite{jacobson} and describe how
our study will go further.

\subsection{\label{wettsec} Review of Wetterich's analysis}

In ref. \cite{wetterich:2002} Wetterich claimed that condition (\ref%
{wettcond}) would hold for any potential, $V^{(w)}(\phi )$, with the
properties $V_{,\phi }(\phi _{c})\gg B_{,\phi }(\phi _{c})\kappa \epsilon
_{c}$. The demonstration was confined to a given background and particular
potential $V^{(w)}(\phi )\propto \exp (-\lambda \phi )$. It was argued that
similar behaviour should be found whenever the potential dominates the
cosmic evolution of the dilaton field i.e. $V_{,\phi }(\phi _{c})\gg
B_{,\phi }(\phi _{c})\kappa \epsilon _{c}$. In what follows we will show
that condition (\ref{wettcond}) also holds in many situations where the
potential does not dominate the cosmic evolution of the dilaton field. It is
our belief, however, that the reasoning given in ref. \cite{wetterich:2002}
is incomplete and does not show that condition (\ref{wettcond}) holds even
under the restrictive conditions specified there. We shall briefly outline
the arguments given in ref, \cite{wetterich:2002} below, and show where we
believe they fail.

Wetterich considered a universe filled with pressureless dust and a
cosmological constant. Under these assumptions we have $T=\epsilon _{dust}$,
the dust density, and:
\begin{equation*}
\square \phi =B_{,\phi }(\phi )\kappa \epsilon _{dust}-V_{,\phi
}^{(w)}\left( \phi \right) ,
\end{equation*}%
where $\epsilon $ is the local matter density and it was assumed that the
potential is
\begin{equation*}
V^{(w)}\left( \phi \right) =\omega e^{-\phi },
\end{equation*}%
where $\omega \sim \mathcal{O}\left( M_{pl}^{2}\right) $. The function $%
B_{,\phi }(\phi )$ represents the coupling of the dilaton field to the dust.
Experimental bounds on the largest allowed violations of the Weak
Equivalence Principle (WEP) that will be created by a dilaton field that
couples to the electromagnetic energy of matter suggest that $\left\vert
B_{,\phi }(\phi )\right\vert <10^{-4}$. Wetterich considered a particularly
simple example of a spherically symmetric massive body superimposed onto the
cosmological background; table \ref{wetttable} shows the local energy
density budget in this model.
\begin{table}[tbp]
\caption{Density distribution in Wetterich's model}
\label{wetttable}
\begin{center}
\begin{tabular}{|c|c|c|c|}
\hline
Region & Range & $\epsilon_{dust}$ & Description \\ \hline
a & $r < R_{s} $ & $\epsilon = \epsilon_{E}$ & local planet \\ \hline
b & $R_{s} < r < r_c$ & $\epsilon = 0$ & intermediate space \\ \hline
c & $r > r_c$ & $\epsilon = \epsilon_c(t)$ & Hubble flow \\ \hline
\end{tabular}%
\end{center}
\end{table}
For $r\gg r_{c}$, $\phi $ takes it cosmological value $\phi _{c}(t)$, which
satisfies:
\begin{equation*}
\ddot{\phi}_{c}+3H\dot{\phi}_{c}=\omega e^{-\phi }+B_{,\phi }(\phi _{c})\kappa
\epsilon _{c}.
\end{equation*}%
Wetterich only considered the case where $\omega e^{-\phi _{c}}\gg B_{,\phi
}(\phi _{c})\kappa \epsilon _{c}$. Assuming that the scale factor $a\propto
t^{n}$, we find:
\begin{equation*}
\phi _{c}(t)=\phi _{0}+2\ln (t/t_{0}),
\end{equation*}%
where $t=t_{0}$ today, and $\phi (t_{0})=\phi _{0}=-\ln(6n+2)/\omega t_{0})\approx 140+\ln \left(
\omega /M_{pl}^{2}\right) $ is the present value of the scalar field. For $%
r\approx R_{s}$, dilaton field in ref. \cite{wetterich:2002} is written as:
\begin{equation*}
\phi =\phi _{c}(t)+\phi _{l}(r,t)+\phi _{e}(r,t),
\end{equation*}%
\noindent where this defines $\phi_{e}$ and $\phi _{l}(r,t)$ is the `local', quasi-static
field configuration satisfying:
\begin{equation*}
\nabla ^{2}\phi _{l}-\mu ^{2}\phi _{l}=-B_{,\phi }\left( \phi _{c}\right)
\kappa \epsilon _{E}\theta \left( R_{s}-r\right) ,
\end{equation*}%
with $\nabla ^{2}$ the 3-D Laplacian and $\theta \left( R_{s}-r\right) $ the
Heaviside function; $\mu ^{2}=V_{,\phi \phi }^{(w)}(\phi _{c})$ and $\mu
^{2}R_{s}^{2}\ll 1$. Near $r=R_{s}$ it was found that $\dot{\phi}_{l}/\dot{%
\phi}_{c}\sim B_{,\phi }(\phi _{c})2M/R_{s}\ll 1$. If $\dot{\phi}_{e}(r,t)/%
\dot{\phi}_{c}(t)\ll 1$ for $r\sim R_{s}$ then the dilaton field will
satisfy condition (\ref{wettcond}). If $|\square \phi _{e}/\square \phi
_{c}|\ll 1$ locally then, as stated in ref. \cite{wetterich:2002} we will
have the required result. Wetterich argues that this is indeed the case in
his paper. However, this does not appear to be the case. Assuming $%
2m/R_{s}\ll 1$, from $r=R_{s}$ spacetime is approximately Minkowski and so:
\begin{eqnarray}
\square \phi _{e} &=&\ddot{\phi}_{e}-\nabla ^{2}\phi _{e}\approx -\mu
^{2}\phi _{e}-\ddot{\phi}_{l}+3H\dot{\phi}_{c}-B_{,\phi }(\phi _{c})\epsilon
_{c}+B_{,\phi \phi }(\phi _{c})\epsilon _{E}\theta (R_{s}-r)\left( \phi
_{l}+\phi _{e}\right) , \\
\square \phi _{c} &=&\ddot{\phi}_{c}=-3H\dot{\phi}_{c}-V_{,\phi }^{(w)}(\phi
_{c}).
\end{eqnarray}%
Now, $\ddot{\phi}_{l}\ll \ddot{\phi}_{c}$ and $\mu ^{2}R_{s}^{2}\ll 1$ and
so, near the surface of our body, the $\mu ^{2}\phi _{e}$ term represents a
negligible correction to the $\phi _{e}$ dynamics. Therefore, if $\square
\phi _{c}\gg \square \phi _{e}$ we must have:
\begin{equation*}
\left( 3H\dot{\phi}_{c}-B_{,\phi }(\phi _{c})\epsilon _{c}+B_{,\phi \phi
}(\phi _{c})\epsilon _{E}\left( \phi _{l}+\phi _{e}\right) \right) \ll \ddot{%
\phi}_{c}.
\end{equation*}%
However, under Wetterich's assumptions that $B_{,\phi \phi }(\phi _{c})\sim
\mathcal{O}\left( B_{,\phi }(\phi _{c})\right) $ and $\epsilon _{E}\sim
10^{30}\epsilon _{c}$, as is the case for the density of the Earth, we find:
\begin{equation*}
\frac{B_{,\phi \phi }(\phi _{c})\epsilon _{E}\phi _{l}}{\ddot{\phi}_{c}}%
\approx 10^{22}\left( \frac{B_{,\phi }(\phi _{c})}{10^{-4}}\right) ^{2}\frac{%
2m}{R_{s}}\approx 10^{13}\left( \frac{B_{,\phi }(\phi _{c})}{10^{-4}}\right)
^{2}\gg 1,
\end{equation*}%
where we have taken $\frac{2m}{R_{s}}\sim 10^{-9}$ in the final deduction.
Hence, we have shown that in general the condition $|\square \phi
_{e}/\square \phi _{c}|\ll 1$ does \emph{not} hold; indeed $\square \phi
_{e}\gg \square \phi _{c}$. This result is opposite to the one found in ref.
\cite{wetterich:2002}. We conclude then that Wetterich's analysis does not
prove that condition (\ref{wettcond}) holds. If we approximate our local
solution by $\phi =\phi _{l}+\phi _{c}$ then we have seen that correction
terms, $\phi _{e}$, to this solution vary on scales much smaller than $1/%
\dot{\phi}_{c}$. As a result we cannot conclude that $\dot{\phi}\approx \dot{%
\phi}_{c}$. In fact, at the epoch $t=t_{0}$, the asymptotic expansion of the
local solution should be correctly written as:
\begin{equation*}
\phi \sim \phi _{0}+\phi _{l}(r,t_{0}),
\end{equation*}%
We shall see later by more detailed methods that, even though the above
analysis fails to show it, that we should expect condition (\ref{wettcond})
to hold for this set-up.

\subsection{\label{jacsec} Jacobson's result}

In ref. \cite{jacobson}, the problem of gravitational memory \cite{mem1} is
considered in the context of Brans-Dicke (varying-$G$) theory. If Newton's
constant, $G$, can and does vary over time and space then one must ask which
value of $G$ is appropriate for use on the horizon of a black hole after it
forms. One motivation was to discover if the black hole possesses a type of
gravitational memory, freezing-in the value of $G$ that existed
cosmologically at the moment when it formed in the early universe, or whether
the value of Newton's `constant' on the horizon changes over time so as to
track its changing cosmological value in the background universe. This has
been investigated by several different methods and found to be a small
effect \cite{mem2} but Jacobson's argument was that if the cosmological
variation in $G$, and the related Brans-Dicke field, $\phi \propto G^{-1}$,
is slow over time-scales of the order of the intrinsic length scale of the
black-hole ($\sim 2GM$) then, at each epoch $t=t_{0}$, one can expand the
cosmological value of $\phi $ as a Taylor series in ($T=t-t_{0}$) and, to a
good approximation drop all $\mathcal{O}(T^{2})$ and higher-order terms, so
\begin{equation*}
\phi _{c}\approx \phi _{1}=\phi _{c}(t_{0})+\dot{\phi}_{c}(t_{0})T.
\end{equation*}%
Jacobson noted that $\phi _{1}$ is a complementary solution to the
Brans-Dicke field conservation equation in an (empty) Schwarzschild
background:
\begin{equation*}
\square _{s}\phi _{1}=0,
\end{equation*}%
\noindent where $\square _{s}$ is the d'Alembertian operator for the
Schwarzschild metric. Therefore $\phi _{1}$ can be added to any known
Schwarzschild-background solution of Brans-Dicke field equations to gain a
new solution. Jacobson took $T$ to be the `curvature' (or Schwarzschild)
time so that the metric is given by:
\begin{equation*}
\mathrm{d}s^{2}=\left( 1-\frac{2m}{R}\right) \mathrm{d}T^{2}-\left( 1-\frac{%
2m}{R}\right) ^{-1}\mathrm{d}R^{2}-R^{2}\{\mathrm{d}\theta ^{2}+\sin
^{2}\theta \mathrm{d}\phi ^{2}\}.
\end{equation*}%
This time coordinate diverges as we move towards the horizon, hence $\phi
_{1}(T)$ also diverges in this limit. The unique static solution of the
dilaton conservation equation, which vanishes as $r\rightarrow \infty$, is given
by:
\begin{equation*}
\phi _{2}=\ln \left( 1-\frac{2m}{r}\right)
\end{equation*}%
Whilst both $\phi _{1}$ and $\phi _{2}$ diverge on the horizon, Jacobson
found that there is a unique linear combination of them that is well-defined
as $r\rightarrow 2m$:
\begin{equation}
\phi _{jac}=\phi _{1}(t)+2m\dot{\phi}_{c}(t_{0})\phi _{2}(r)=\dot{\phi}%
_{c}\left( v-r-2m\ln (r/2m)\right)  \label{phijac}
\end{equation}%
\noindent where $v=t+r+2m\ln (r/2m-1)$ is the advanced time coordinate. By
construction, it can then be shown that there exists a unique solution for $%
\phi $ in the Schwarzschild background that is non-singular on the horizon
and has $\phi _{jac}(r=\infty ,t)\approx \phi _{c}(t)$ for the particular
case where $\phi _{c}(t)\propto t\propto G^{-1}$. If the approximation used
is valid, then this suggests that the value of $G$ on the horizon lags
slightly behind the cosmological value, but that over time-scales much
larger than $m$ there is no gravitational memory when $G\propto t^{-1}$
falls in this extreme fashion.

Jacobson's result assumes that the cosmological region can be taken to be
infinitly far away, so that the entire spacetime is Schwarzschild; in reality
the cosmological matter will become gravitational dominant over the
black-hole at some finite-distance. Another issue with Jacobson's method is
that, since $T$ diverges on the horizon, the expansion of $\phi _{c}$ as a
Taylor series in small $T$ will not be valid near the horizon. In realistic
models the space surrounding the black hole will also not be totally empty,
indeed there will be an accretion disk surrounding the black-hole. The
time-scale for matter to fall into the black-hole is order$\sqrt{r^{3}/2m}$
and this is relatively short compared with the time-scale over which
cosmological expansion occurs. Accretion of matter into the black-hole,
therefore, might well result in a significant difference in the
time-variation of $G$ close to the horizon or cosmologically. In this paper
we will consider a more realistic embedding of a Schwarzschild mass in an
expanding universe; and build our asymptotic expansions in such a way that
they are well-defined on the horizon. We will conclude that Jacobson's
result \cite{jacobson} (see also refs. \cite{mem2, harada} for similar
conclusions arrived at by different methods) does indeed give the correction
behaviour of $\phi $ whenever the energy-density in the region surrounding
the black-hole is low enough and for more general time-variations of $G(t)$
in the background universe.

\section{Matched Asymptotic Expansions}

The method of matched asymptotic expansions that we will use in this paper
was first developed to solve systems of PDEs that involve multiple length
scales. It is often used in the field of fluid dynamics to study systems
where there is thin boundary layer, inside which the length scale of
variation is much smaller than outside it. It also has applications in the
study of slender bodies, with widths much smaller than their lengths. Other
problems with multiple length scales include the interaction of greatly
separated particles and the evaluation of the influence of a slowly changing
background field on the dynamics of a small body. The books by Cole, \cite%
{Cole}, and Hinch, \cite{Hinch}, provide a more detailed treatment of this
subject. In this section we will briefly introduce the method and give some
simple examples of its applications in order to fix ideas.

\subsection{Asymptotic expansions}

Critical to this method is the requirement that the approximations we will
work with are \emph{asymptotic} expansions rather than convergent
(Taylor-like) power-series. In general, an asymptotic expansion will not be
convergent. Thus, we define asymptotic approximations and expansions as
follows:

\begin{defn}
\label{defasymp} $\sum^{M}f_{n}(\delta )$ is an \emph{asymptotic
approximation} to $f(\delta )$ as $\delta \rightarrow 0$, if for each $M\leq
N$ the remainder term is much smaller than the last included term:
\begin{equation*}
\frac{f(\delta )-\sum^{M}f_{n}(\delta )}{f_{M}(\delta )}\rightarrow 0\text{ }%
\mathrm{as}\;\delta \rightarrow 0.
\end{equation*}%
One then writes:
\begin{equation*}
f(\delta )\sim \sum^{N}f(\delta )\text{ }\mathrm{as}\;\delta \rightarrow 0.
\end{equation*}
\end{defn}

\begin{defn}
If definition \ref{defasymp} holds, in principle, for all $N$, i.e. we can
take $N=\infty$, then the we deem the approximation to be an \emph{%
asymptotic expansion} of $f(\delta)$:
\begin{eqnarray}  \label{asympeq.}
&f(\delta) \sim \sum^{\infty} f_{n}(\delta)&\mathrm{as} \; \delta
\rightarrow 0.
\end{eqnarray}
\end{defn}

In many cases $f_{n}(\delta )\propto \left( \delta ^{p}\right) ^{n}$ for
some constant $p$ and we will have an \emph{asymptotic power series}. It is
also quite common to find $f_{i}(\delta )\propto \delta ^{j}\ln (\delta )$,
for some $i$ and $j$. The sum in eq. (\ref{asympeq.}) is a formal one, and
it is not required that it does converge. The property required by
definition \ref{defasymp} is, however, more useful that convergence in many
cases; it ensures that one only needs the first few terms of the expansion
to create a good numerical approximation to $f(\delta )$.

\subsection{Singular problems and ones with multiple scales}

Consider a differential operator $\mathcal{L}_{x}(\delta )$ which defines
some function $f(x,\delta )$ by $\mathcal{L}_{x}(\delta )f(x,\delta )=0$. We
can attempt to solve this equation by making an asymptotic expansion of $%
f(x,\delta )$ and solving the resultant equation order-by-order in $\delta $%
:
\begin{equation*}
f(x,\delta )\sim \sum^{\infty }f_{n}(x)\gamma _{n}(\delta )\mathrm{as}%
\;\delta \rightarrow 0,
\end{equation*}%
with $x$ fixed. A 'singular' problem is one where the above asymptotic
expansion is not \emph{uniformly valid}, i.e. it breaks down for certain
ranges of $x$, typically $x=\mathcal{O}\left( \delta \right) $ or $x=%
\mathcal{O}\left( 1/\delta \right) $. Singular behaviour such as this can be
divided into two distinct classes. In the first, the highest derivative in $%
\mathcal{L}_{x}$ is multiplied by some power of $\delta $ and so can be
ignored everywhere apart from in those regions where the variation in $%
f(x,\delta )$ is fast enough to ensure that the highest derivative make a
significant contribution. In the second class, the problem has more than one
intrinsic length (or time) scale, one much larger than the other. The
application of this method to physical problems generally tends to fall into
this latter class. In both cases one proceeds by constructing two (or more)
asymptotic approximations to the solutions which are valid for different
ranges of $x$, e.g. for $x\sim \mathcal{O}(1)$ and $x/\delta =\xi \sim
\mathcal{O}(1)$, with
\begin{eqnarray}
&f(x,\delta )\sim \sum_{n=0}^{Q}f_{n}(x)\delta _{n}\;&\mathrm{as}\;\delta
\rightarrow 0,\;x\;\mathrm{fixed},  \label{outer} \\
&f(x,\delta )\sim \sum_{n=0}^{P}g_{n}(\xi )\delta _{n}\;&\mathrm{as}\;\delta
\rightarrow 0,\;\xi =x/\delta \;\mathrm{fixed},  \label{innersoln}
\end{eqnarray}%
and solving $\mathcal{L}_{x}(\delta )f(x,\delta )$ order by order in $\delta
$ for both expansions w.r.t. to some boundary conditions. We will call
expansion (\ref{outer}) the \emph{outer solution}, and (\ref{innersoln}) the
\emph{inner solution}. The inner expansion is not uniformly valid in the
region $\xi =\mathcal{O}(1/\delta )$, as the outer one is not valid where $x=%
\mathcal{O}(\delta )$. Because of these restrictions on the size of $x$, we
will only be able to apply a subset of the boundary conditions to each
expansion; in general we will, therefore, be left with unknown coefficients
in our asymptotic approximations. This ambiguity can be lifted by matching
the inner and outer solutions in an intermediate region where they are both
valid.

\subsection{\label{matchsec1} Matching}

We can match the inner and outer solutions together if there exists some
intermediate range of $x$ where both (\ref{innersoln}) and (\ref{outer}) are
valid asymptotic approximations to $f(x,\delta )$. By the uniqueness
properties of asymptotic expansions, see \cite{Hinch} for a proof, if they
are both valid in some region then the two expansions must be equal. We take
$x$ to scale as some intermediate function, $\eta ({\delta )}$, with
magnitude between $1$ and $\delta $, e.g. $\eta ({\delta )}=\delta ^{\alpha
} $ where $0<\alpha <1$, and define a new $X$ coordinate appropriate for the
intermediate region by
\begin{equation*}
X=x/\eta (\delta )=\left( \frac{\delta }{\eta (\delta )}\right) \xi .
\end{equation*}%
We then write both the inner and outer approximations in terms of $X$ and,
keeping $X$ fixed, make an asymptotic expansion of each of them in the limit
$\delta \rightarrow 0$; this is called the \emph{intermediate} limit. We
must be careful to neglect any terms in the approximations, that we find in
this way, which would be smaller than the first excluded term; equality is
then demanded between the remaining terms:
\begin{eqnarray}
&\sum_{n=0}^{Q}f_{n}(x(X,\eta ))\delta _{n}\sim
\sum_{n=0}^{S}h_{n}^{(out)}(X)\gamma _{n}(\delta )&\mathrm{as}\;\delta
\rightarrow 0;\;X\;\mathrm{fixed},  \notag  \label{match} \\
&\sum_{n=0}^{P}g_{n}(\xi (X,\eta ))\delta _{n}\sim
\sum_{n=0}^{S}h_{n}^{(in)}(X)\gamma _{n}(\delta )&\mathrm{as}\;\delta
\rightarrow 0;\;X\;\mathrm{fixed},  \notag \\
&h_{n}^{(out)}(X)=h_{n}^{(in)}(X)&\mathrm{for}\;0\leq n\leq S.
\end{eqnarray}%
Usually, the precise form we choose for $\eta (\delta )$ is unimportant and
the matching can be done regardless; there are some cases, however, where we
must be more careful (see Hinch \cite{Hinch} for more).

\subsection{Simple Examples}

\subsubsection{\label{eg1} Example 1: Simple matching}

We will start with a boundary layer example where the exact solution is
known. Consider $y(x,\delta )$ which satisfies:
\begin{equation*}
2\delta y^{\prime \prime }+(1+2\delta )y^{\prime }+y=0,
\end{equation*}%
in $0\leq x\leq 1$, with boundary conditions $y=0$ at $x=0$ and $y=e^{-1}$
at $x=1$ and where ${}^{\prime} = \diff/\diff x$. The exact solution is:
\begin{equation}
y(x,\delta )=e^{-1}\left( \frac{e^{-x}-e^{-x/2\delta }}{e^{-1}-e^{-1/2\delta
}}\right) .  \label{exact}
\end{equation}%
If we apply the method of matched asymptotic expansions to this, we would
find that the outer approximation is given (to all orders) by:
\begin{equation*}
y(x,\delta )\sim e^{-x}\neq 0\;\mathrm{fixed}.
\end{equation*}%
We have been able to apply the boundary condition at $x=1$; the $x=0$
boundary condition cannot be reached given our assumptions about the size of
$x$. For the inner approximation we keep $\xi =x/\delta $ fixed. In terms of
$\xi $ our differential equation reads:
\begin{equation*}
2y_{,\xi \xi }+(1+2\delta )y_{,\xi }+\delta y=0.
\end{equation*}%
The inner solution is found to be:
\begin{equation*}
y(x,\delta )\sim a_{0}\left( 1-e^{-\xi /2}\right) +\delta \left( a_{1}\left(
1-e^{-\xi /2}\right) -a_{0}\xi \right) +\delta ^{2}\left( a_{2}\left(
1-e^{-\xi /2}\right) -a_{1}\xi +\tfrac{1}{2}a_{0}\xi ^{2}\right) +\mathcal{O}%
\left( \delta ^{3}\right) ,
\end{equation*}%
with $\xi =x/\delta $ held fixed. Again we can only apply one boundary
condition, this time the one at $x=0$. The boundary condition at $x=1$ is
beyond our reach in this approximation.
\begin{figure}[tbh]
\begin{center}
\includegraphics[scale=0.75]{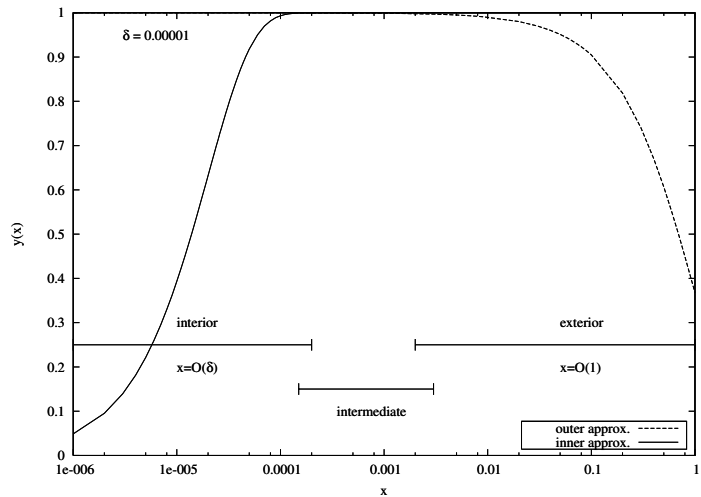}
\end{center}
\caption{The leading order, inner and outer approximations to the solution
of problem given in section \ref{eg1}, with $\protect\delta =0.00001$. The exact
solution is not visible on this plot since it lies underneath
one or other of the approximations everywhere.}
\label{fig1}
\end{figure}
In this problem the choice of the intermediate scale turns out to be
unimportant; we take $\eta =\delta^{1/2}$ and define $X=\delta
^{-1/2}x=\delta^{1/2}\xi $. The intermediate limit of the inner
approximation is:
\begin{equation*}
\mathrm{inner}\;\mathrm{approx.}\sim a_{0}-\delta ^{1/2}a_{0}X+\delta \left(
a_{1}+\tfrac{1}{2}a_{0}X^{2}\right) +\mathcal{O}\left( \delta^{3/2}\right) .
\end{equation*}%
The first excluded term is order $\delta^{3/2}$; we have neglected all
terms of, or below, this order including the exponentially small $e^{-X/2\delta^{1/2}}$ terms. Expanding the outer approximation we have:
\begin{equation*}
\mathrm{outer}\;\mathrm{approx.}\sim 1-\delta ^{1/2}X+\delta \tfrac{1}{2}%
X^{2}+\mathcal{O}\left( \delta ^{3/2}\right) .
\end{equation*}%
By our matching criteria, eq. \ref{match}, we must have:
\begin{eqnarray}
a_{0} &=&1,  \notag \\
a_{1} &=&0.  \notag
\end{eqnarray}%
If we were to perform this process to all orders we would find that all $%
a_{i}=0$ for $i\geq 1$. The fully specified inner approximation is therefore
\begin{equation*}
y(x,\delta )\sim -e^{-\xi/2}+\sum_{n=0}^{\infty}\delta^{n}(-1)^{n}\xi^{n}.
\end{equation*}%
This is precisely what we would have found by performing a Taylor series
expansion of the exact solution, (\ref{exact}), in the inner limit and then
dropped all exponentially small terms i.e. $e^{-1/\delta}$. By performing
the matching we have been able to lift the ambiguity in the coefficients, $%
a_{i}$, and fully specify the inner approximation.

\subsubsection{\label{scalexp} Example 2: Scalar field in an expanding
universe}

Consider a spherically-symmetric, scalar field, $\phi (r,t)$, like the
dilaton, in a flat FRW cosmology with metric
\begin{equation*}
\mathrm{d}s^{2}=\mathrm{d}t^{2}-a^{2}(t)\left( \mathrm{d}r^{2}+r^{2}\left(
\mathrm{d}\theta ^{2}+\sin ^{2}\theta \mathrm{d}\phi ^{2}\right) \right) .
\end{equation*}%
To make contact with our problem we will assume that the only source term in
the $\phi $ evolution equation is homogeneous and proportional to the
background dust density, so
\begin{equation*}
\square \phi =\frac{(a^{3}(t)\dot{\phi})^{\cdot }}{a^{3}(t)}-\frac{(r\phi
)^{\prime \prime }}{a^{2}(t)r}=\frac{\lambda H_{0}^{2}}{a^{3}(t)}
\end{equation*}%
where ${}^{\cdot} = \partial/\partial t$ and ${}^{\prime} = \partial / \partial r$ and with $H_{0}$ the Hubble parameter at some arbitrary time $t=t_{0}$ and $%
\lambda $ is a constant parameter of the theory. We simplify the problem
further by considering only the matter era, where $a(t)\propto t^{2/3}$. As
boundary conditions we take $\phi \rightarrow \phi _{0}(t)$ as $r\rightarrow
\infty $ and $\phi ^{\prime }/a(t)|_{r=r_{0}/a(t)}=\mu 2m/r_{0}^{2}=const$; $%
\mu $ is a dimensionless constant, and $m$ is a length scale such that $%
\delta =2mH_{0}\ll 1$. We assume that $r_{0}/a(t_{0})\sim \mathcal{O}(2m)$.
We could visualise this problem arising from the presence of a massive body
in the region $a(t)r<r_{0}$ that creates a spatial gradient in $\phi $. Near
$a(t)r=r_{0}$ we have $\phi ^{\prime }\gg \dot{\phi}$. We take our problem
to be similar to one we will consider later and determine the relation
between $\dot{\phi}_{0}$ and $\dot{\phi}(r_{0}/a(t),t)$.

The small parameter in this problem is $\delta =2mH_{0}$. We work at the
epoch when $t=t_{0}$; for simplicity we take $a(t_{0})=1$. In the inner
approximation we define $2m\xi =a(t)r$, and $\xi _{0}:=r_{0}/2m\sim \mathcal{%
O}(1)$. We assume that the inner solution is quasi-static, so it depends on
time only through the slowly increasing cosmological $t$, rather than $%
(t-t_{0})/2m$. We define a dimensionless time coordinate by $\tau =H_{0}t$, $%
\tau _{0}=H_{0}t_{0}=2/3$. In this interior region $\phi =\phi _{I}(\xi
,\tau ;\delta )$ which satisfies

\begin{equation*}
\frac{1}{\xi }\partial _{\xi }^{2}\left( \xi \phi _{I}\right) =\delta ^{2}%
\left[ \frac{1}{\tau ^{2}}\partial _{\tau }\left( \tau ^{2}\partial _{\tau
}\phi _{I}\right) -\frac{4\lambda }{9\tau ^{2}}+\frac{2}{3\tau ^{2}}\xi
\partial _{\xi }\phi _{I}+\frac{4}{9\tau ^{2}}\left( \xi \partial _{\xi
}\right) ^{2}\phi _{I}+\frac{4}{3\tau }\xi \partial _{\xi }\partial _{\tau
}\phi _{I}\right] .
\end{equation*}%
\noindent Solving this to order $\mathcal{O}\left( \delta ^{2}\right) $ we
find:
\begin{equation*}
\phi _{I}(\tau ,\xi )\sim \phi _{0}^{I}\left( \tau \right) -\frac{\mu }{\xi }%
+\delta ^{2}\left( \phi _{1}^{I}\left( \tau \right) +\frac{\mu }{9\tau ^{2}}%
\left( \xi +\frac{\xi _{0}^{2}}{\xi }\right) +\frac{g(\tau )\left( \xi
^{3}-2\xi _{0}^{3}\right) }{6\xi }\right) +\mathcal{O}\left( \delta
^{4}\right)
\end{equation*}%
where $\phi _{0}^{I}\left( \tau \right) $ and $\phi _{1}^{I}\left( \tau
\right) $ are `constants' of integration, to be determined via the matching
procedure. $g(\tau )=\frac{1}{\tau ^{2}}\partial _{\tau }\left( \tau
^{2}\partial _{\tau }\phi _{0}^{I}\right) -\frac{4\lambda }{9\tau ^{2}}$. In
the exterior we define $\rho =H_{0}r$ to be our dimensionless radial
coordinate, and $\phi =\phi _{E}$. Our boundary conditions determine the
leading-order exterior term to be $\phi _{0}(\tau )$ ($=\phi _{0}(t)$ with
some abuse of notation). The sub-leading order terms, $\delta ^{i}\phi
_{E}^{(i)}$ are then given by $\square \phi _{E}^{(i)}=0$. So, to order $%
\delta $,
\begin{equation*}
\phi _{E}(\tau ,\rho )\sim \phi _{0}\left( \tau \right) +\delta \frac{%
\int_{-\infty }^{\infty }\mathrm{d}\gamma T_{\gamma }(\tau )X_{\gamma }(\rho
)}{\rho }+\mathcal{O}(\delta ^{2})
\end{equation*}%
\noindent where
\begin{eqnarray}
&X_{\gamma }(\rho )=A(\beta )\cos \left( \beta \rho \right) +B(\beta )\sin
\left( \beta \rho \right) &\mathrm{where}\;\gamma =-\beta ^{2}<0,
\label{taueq.} \\
&X_{\gamma }(\rho )=A(0)&\mathrm{where}\;\gamma =0,  \notag \\
&X_{\gamma }(\rho )=C(\alpha )e^{-\alpha \rho }&\mathrm{where}\;\gamma
=\alpha ^{2}>0,  \notag \\
&\tau ^{2}T_{\gamma }(\tau )_{,\tau \tau }+2\tau T_{\gamma }(\tau )_{,\tau
}=\gamma \tau ^{2/3}T_{\gamma }(\tau ),&
\end{eqnarray}%
and $A(\beta )$, $B(\beta )$, $C(\alpha )$ are all to be determined by the
matching procedure. As in the previous example, the precise position of the
intermediate region is not important. We choose an intermediate coordinate $%
z=\delta ^{-1/2}\rho =\delta ^{1/2}\xi $ and take the intermediate limit of
both the interior and exterior approximations. By equating our two
approximations in the intermediate region we find $B(\beta )=C(\beta )=0$
from the $\mathcal{O}\left( \delta \right) $ terms and
\begin{eqnarray}
&\mathcal{O}\left( 1\right) :\;&\phi _{0}^{I}(\tau )=\phi _{0}(\tau
)\rightarrow g(\tau )=0,   \\
&\mathcal{O}\left( \delta ^{1/2}\right) :\;&\int_{0}^{\infty }\mathrm{d}%
\beta A(\beta )T_{-\beta ^{2}}\left( \tau \right) =\frac{\mu }{\tau ^{2/3}}
\label{matchex1} \\
&\mathcal{O}\left( \delta ^{3/2}\right) :\;&\int_{0}^{\infty }\mathrm{d}%
\beta \beta ^{2}A(\beta )T_{-\beta ^{2}}\left( \tau \right) =\frac{2\mu }{%
9\tau ^{4/3}}.\label{matchex2}
\end{eqnarray}%
By differentiating twice and applying (\ref{taueq.}) we can check that
conditions (\ref{matchex1}) and (\ref{matchex2}) can be simultaneous
satisfied for some choice of $A(\beta )$. The $T_{\gamma }\left( \tau
\right) $ can be made orthonormal w.r.t. to some inner product and so eq. (%
\ref{matchex1}) can, in principle, be inverted to find $A(\beta )$. We omit
this step, however, since we are mostly concerned with the effect of the
exterior on the behaviour of $\phi $ in the interior rather than vice versa.
By finding the interior solution to $\mathcal{O}\left( \delta ^{4}\right) $
and the exterior to $\mathcal{O}\left( \delta ^{2}\right) $ we can show that
$\phi _{1}^{I}\left( \tau \right) =0$.; we have now fully specified the
interior solution to $\mathcal{O}\left( \delta ^{2}\right) $:
\begin{equation*}
\phi _{I}\left( t,R=a(t)r\right) \sim \phi _{0}\left( t\right) -\frac{2m\mu
}{R}+\frac{4m\mu }{9t^{2}}\left( R+\frac{r_{0}^{2}}{R}\right) +\mathcal{O}%
\left( \delta ^{4}\right)
\end{equation*}%
We have that:
\begin{equation*}
\frac{\dot{\phi}_{I}|_{ar=r_{0}}}{\dot{\phi}_{0}}-1\approx \frac{16m\mu r_{0}%
}{9t^{3}\dot{\phi}_{0}}
\end{equation*}%
and $\dot{\phi}_{0}\sim 4\lambda /9t$ for large $t$. Therefore we have shown
that for the time variation of $\phi $ at $ar=r_{0}$ to track the
cosmological time variation we need
\begin{equation*}
\lambda \gg \frac{2mr_{0}}{t^{2}}=\mathcal{O}\left( \delta ^{2}\right)
\end{equation*}%
It is clear that whatever the value of $\lambda $, the rate of time
variation in $\phi $ will tend to homogeneity as $t\rightarrow \infty $.
This example is a greatly simplified version of the problem we will consider
in the rest of this paper.

\subsection{Application to General Relativity}

The use of matched asymptotic expansions in general relativity was pioneered
by Burke and Thorne \cite{burkethorne}, Burke \cite{burke}, and D'Eath \cite%
{Death1, Death2} in the 1970s. These authors used them to the study how the
laws of motion of a test body were affected by the background spacetime. We
shall now outline how matched asymptotic expansions are applied in general
relativity.

We assume that the universe

\begin{equation*}
\mathcal{C}=\left( \mathcal{M},g_{ab},T^{ab},\phi \right) ,
\end{equation*}%
with $T_{ab}$ is the energy-momentum tensor and $\phi $ the dilaton, can be
viewed as a background cosmology,

\begin{equation*}
\mathcal{C}_{0}=\left( \mathcal{M}_{0},g_{ab}^{0},T_{0}^{ab},\phi
_{0}\right) ,
\end{equation*}
onto which some localised, interior configuration has been superimposed in a
non-linear fashion. In what follows, for simplicity, we will require the
interior configuration to be static in some coordinate system. We demand
that the `size' of the interior region be given by a single parameter $m$;
and that as $m\rightarrow 0$, with $\left\{ x^{\mu }\right\} $ fixed, the
interior region disappears and $\mathcal{C}\rightarrow \mathcal{C}_{0}$. We
shall formalise this statement later.

\subsubsection{Length scales}

The length scale of the interior region, as defined by its curvature
invariants is denoted by $L_{I}\left( m\right) $, with $\ L_{I}\left(
m\right) \rightarrow 0$ as $m\rightarrow 0$. The length scale of the
background (exterior) region is written $L_{E}$. For the asymptotic
expansion method to be viable we need $\delta =L_{I}/L_{E}\ll 1$. The effect
of the interior on the background configuration can then be treated as a
linear perturbation to $\mathcal{C}_{0}$, with $\delta $ playing the rôle of
a small parameter. We can similarly treat the effect of the background on
the interior as a linear perturbation.

\subsubsection{Five-dimensional manifold}

\begin{figure}[htb!]
\begin{center}
\includegraphics[scale=0.50]{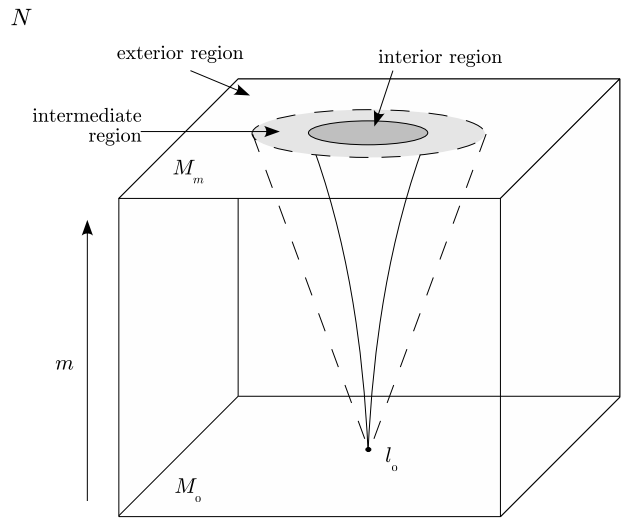}
\end{center}
\caption{The 5-D manifold $\mathcal{N}$ is built up from spacetimes $(%
\mathcal{M}_{m}, g_{ab}(m))$ for $m \in [0,m_{max})$.}
\label{fig2}
\end{figure}

For each $m$, in some interval $[0,m_{max})$, we write the global
configuration as $\mathcal{C}_{m}=\left( \mathcal{M}%
_{m},g_{ab}(m),T^{ab}(m),\phi (m)\right) $. Following Geroch \cite{Geroch}
and D'Eath \cite{Death1}, we consider a five-dimensional manifold with
boundary, $\mathcal{N}$, that is built up from spacetimes $(\mathcal{M}%
_{m},g_{ab}(m))$ for $m\in \lbrack 0,m_{max})$; $g_{ab}(0)=g_{ab}^{(0)}$. As
D'Eath noted we should properly exclude from $\mathcal{M}_{0}$ a smooth
time-like world line, $l_{0}$, that corresponds to the `position' of the
interior region; this is illustrated in figure \ref{fig2}. We require that
the contravariant metrics $g^{ab}(m)$ define a smooth tensor field on $%
\mathcal{N}$; in addition, we require that the dilaton, $\phi (m)$, defines
a smooth scalar field on $\mathcal{N}$. These conditions are required by our
assumption that the interior region has only a small perturbing effect over
length scales $>>L_{I}(m)$.

In an open subset on $\mathcal{N}$, we choose a coordinate chart $%
(t,r,\theta ,\phi ,m)$ such that the aforementioned world line $l_{0}$ is
given by $(r=0,m=0)$. In the limit $(tL_{E}^{-1},rL_{E}^{-1},\theta ,\phi
)\rightarrow $ constants and $m\rightarrow 0$ we can give $g_{ab}(m)$, $%
T^{ab}(m)$ and $\phi (m)$ as asymptotic expansions about $g_{ab}^{(0)}$, $%
T_{(0)}^{ab}$ and $\phi _{0},$ respectively. This is the \emph{exterior
approximation}, and is appropriate for considering the perturbing effect
that the interior has on the background universe.

For each epoch $t=t_{0}$, we shall also define an \emph{interior
approximation}. For this, we take the asymptotic expansions of $g_{ab}(m)$, $%
T^{ab}(m)$ and $\phi (m)$ in the limit $\left(
(t-t_{0})L_{I}^{-1}(m),rL_{I}^{-1}(m),\theta ,\phi \right) \rightarrow $
constants, and $m\rightarrow 0$. This approximation is appropriate for
considering the perturbations produced in the interior region by the
background cosmology. This is the problem that we are most interested in. We
shall assume that we know the leading-order configuration in this limit and
in what follows we take it to be a Schwarzschild metric with a quasi-static
dilaton field.

\subsubsection{Intermediate region and matching}

We match the interior and exterior approximations in some intermediate
region, where $\left( (t-t_{0})L_{int}^{-1},rL_{int}^{-1}\right) \rightarrow
const$ as $m\rightarrow 0$, and where $L_{int}=L_{E}\delta ^{1-\alpha
}=L_{I}\delta ^{-\alpha }$, with $0<\alpha <1$. In many cases the precise
value of $\alpha $ is not important (see section \ref{matchsec1}). Following
D'Eath, \cite{Death1}, we assume that all the functions in our
approximations are sufficiently well-behaved in this region so as to admit a
power-series expansion in the radial coordinate; given this we shall not
need to examine this region explicitly. A typical term in the interior
expansion will, in this matched region, look like $\delta ^{i}\xi
^{j}f_{ij}(t_{0};L_{int}^{-1}(t-t_{0}))$ where $\xi =rL_{I}^{-1}$ and, $i$
and $j$ are rational numbers; $i\geq 0$. In the exterior region, a typical
term will look like $\delta ^{k}\rho ^{l}g_{kl}(t_{0};L_{int}^{-1}(t-t_{0}))$%
, again with $k$ and $l$ rational and $k\geq 0$ and $\rho =L_{E}^{-1}r$.
Matching will require that $g_{(k+1)l}=f_{ij}$. When we consider the
Tolman-Bondi class of background metrics we will modify this procedure
slightly. For these metrics there are two choices for radial coordinate:

\begin{itemize}
\item the physical radial coordinate $R$ defined by surfaces of $(t,R)=%
\mathrm{const}$ with surface area $4\pi R^{2}$,

\item the radial coordinate which is constant on the world-lines of dust
particles, $r$. There is an ambiguity in this definition of $r$ in that the
defining property is also satisfied by any arbitrary function $h(r)$ of $r$.
We lift this ambiguity by demanding that $R=r$ at $t=t_{0}$. At later times
the Einstein equations give us $R=R(r,t)$.
\end{itemize}

The functions in our interior expansion will generally be quasi-static when
written as functions of $R$ and $r$. We will choose to expand the interior
and exterior approximations as power series in the physical radius, $R$,
when taking the intermediate limit.

\section{Geometrical Set-up}

\subsection{General Picture}

We shall assume that the dilaton field is only weakly coupled to gravity,
and so its energy density has a negligible effect on the background
spacetime geometry. If we ignore this back-reaction then we can, rather than
solving the full set of Einstein-matter-dilaton equations, simply consider
the dilaton field's evolution on a fixed background with given matter
density. We will take the background to be some known exact solution to
Einstein's equations with matter possessing the following properties:

\begin{itemize}
\item The metric is approximately Schwarzschild, with mass $m$, inside some
closed region of spacetime outside a surface at $r=R_{s}$. The metric for $%
r<R_{s}$ is left unspecified.

\item Asymptotically, the metric must approach FRW and the whole spacetime
should tend to pure FRW in the limit $m\rightarrow 0$.

\item As the local inhomogeneous energy density of asymptotically FRW
spacetimes tends to zero, the spacetime metric exterior to $r=R_{s}$ must
tend to a Schwarzschild metric with mass $m$ .
\end{itemize}

\begin{figure}[tbh]
\begin{center}
\includegraphics[scale=0.75]{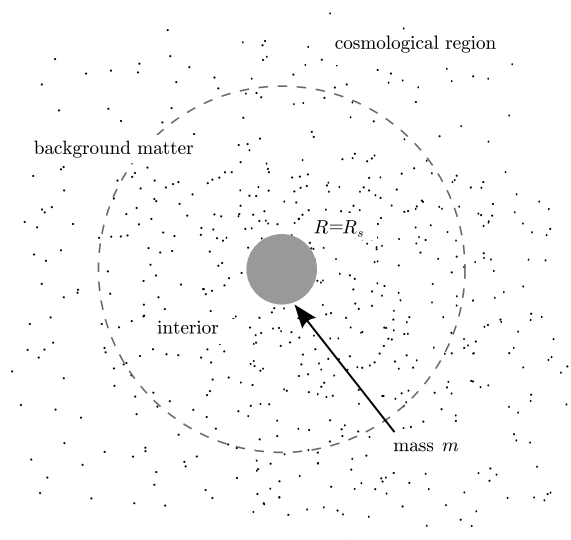}
\end{center}
\caption{We are concerned with the evolution of the dilaton field, $\protect%
\phi $, in spacetimes where a mass $m$, of radius $R_{s}$, has been embedded
into the cosmological fluid.}
\label{fig3}
\end{figure}
This list of requirements is far from exhaustive, and they shall be
re-expressed in more rigorous fashion later. In the local, approximately
Schwarzschild, region the intrinsic length scale, $L_{I}$, of a sphere
centred on the Schwarzschild mass with surface area $4\pi R_{s}^{2}$, is
given by considering the curvature invariant:
\begin{equation}
L_{I}=\left( \tfrac{1}{12}R_{abcd}R^{abcd}\right) ^{-1/4}=\frac{R_{s}^{3/2}}{%
\left( 2m\right) ^{1/2}}.  \label{invar}
\end{equation}%
In the asymptotically FRW region, the intrinsic length scale is proportional
to the inverse root of the local energy density: $1/\sqrt{\kappa \epsilon
+\Lambda }$. We shall assume that, in line with current observations, that
the FRW region is approximately flat, and so we set our exterior length
scale appropriate for the epoch at $t=t_{0}$ is the inverse Hubble parameter
at that time:
\begin{equation*}
L_{E}=1/H_{0}.
\end{equation*}%
Our small parameter is then defined to be

\begin{equation*}
\delta =L_{I}/L_{E}.
\end{equation*}

Formally we require that, as $\delta \rightarrow 0,$ our choice of spacetime
background should be FRW at zeroth-order in the exterior limit and
Schwarzschild to lowest order in the interior limit. In section \ref%
{secvalid} we will say more about what is required of the background for
this matching procedure to be valid. In this paper we shall, in addition to
the criteria given above, restrict ourselves to the subcase where the
spacetime is spherically symmetric. All spherically-symmetric solutions to
Einstein's equations with matter, where the matter is pressureless dust and
a cosmological constant and the flow lines of the matter particles are
geodesic, fall into the Tolman-Bondi class of metrics (for a review of these
and other inhomogeneous spherically symmetric metrics see ref \cite{kr}).
This class of possible Tolman-Bondi spacetimes is parametrised by two
arbitrary functions of one spatial variable, $r$. We will only look at the
flat, Tolman-Bondi models with non-simultaneous initial singularities (which
where rediscovered by Gautreau in \cite{Gautreau:1984, kr}), and the
non-flat Tolman-Bondi models with simultaneous initial singularities. These
two subcases are fully specified by prescribing only a single function: the
matter density on some initial space-like hypersurface. Another metric that
has often been used to study the effect of the universe's expansion on solar
system dynamics is the McVittie metric \cite{mcv, kr}. In the exterior limit
the McVittie metric asymptotes to a dust + $\Lambda $ FRW cosmology. For
radii $r$ where $r\gg 2m$, $r\ll H_{0}^{-1}$, the McVittie metric looks like
Schwarzschild spacetime; the horizon in the McVittie metric possesses a
curvature singularity however, and it cannot therefore be used to model a
black hole in an expanding spacetime.\emph{\ }

\subsection{Case I: The McVittie background}

The earliest studies of the gravitational field produced by a
spherically-symmetric, massive body in an expanding universe were based upon
McVittie's solution to Einstein's equations \cite{mcv, kr}. McVittie's
solution is a superposition of the Schwarzschild metric and a FRW
background. In isotropic coordinates the metric, is
\begin{equation}
\mathrm{d}s^{2}=\left[ \frac{1-\mu (t,r)}{1+\mu (t,r)}\right] ^{2}dt^{2}-%
\frac{[1+\mu (t,r)]^{4}}{(1+\tfrac{1}{4}kr^{2})^{2}}a^{2}(t)\left[ \mathrm{d}%
r^{2}+r^{2}\{\mathrm{d}\theta ^{2}+\sin ^{2}\theta \mathrm{d}\phi ^{2}\}%
\right] ,  \label{mcv1}
\end{equation}%
where
\begin{equation}
\mu (t,r)=\frac{m}{2ra(t)}\left( 1+\tfrac{1}{4}kr^{2}\right) ^{1/2},
\label{mcv2}
\end{equation}%
and $m$ and $k$ are arbitrary constants. In the limit $a=1$, $m$ is the
Schwarzschild mass. $k$ is the curvature of the surfaces $(t,r)=\mathrm{const%
}$ in the $m=0$ (i.e. FRW) limit. As a model of a black hole in an expanding
universe it has the distinct disadvantage that the `horizon' is a naked
curvature singularity. This defect aside, the McVittie metric is believed to be a good
approximation to the exterior metric of a massive spherical body with a
physical radius much larger than its Schwarzschild radius.

In the flat ($k=0$) case, the local energy density depends only on time and
is the same as the FRW energy density to which the solution matches smoothly
as $r\rightarrow \infty $. The pressure, however, is not the same as the FRW
pressure. Apart from the vacuum-energy dominated case (i.e. where $\epsilon
=-P$), it is not possible to prescribe a dust, $P=0$, or any other,
non-vacuum, barotropic, $P=P(\epsilon )$, equation of state to hold apart
from in the asymptotic FRW limit.

\subsubsection{Exterior Metric}

We shall work in the $(r,t)$ coordinates introduced above. In the exterior $%
\mathrm{d}r\sim H_{0}^{-1}\sim \mathrm{d}t$. We define dimensionless
exterior coordinates $\tau =H_{0}t$ and $\rho =H_{0}$. Then, with $\delta
=L_{I}/L_{E}=H_{0}R_{s}^{3/2}(2m)^{1/2}\ll 1$ we have:
\begin{eqnarray}
\mathrm{d}s_{ext}^{2} &\sim &H_{0}^{-2}\left\{ g_{ab}^{(0)}+\delta
g_{ab}^{(1)}+\mathcal{O}\left( \delta ^{2}\right) \right\} \mathrm{d}x^{a}%
\mathrm{d}x^{b} \\
&=&H_{0}^{-2}\left\{ \left[ \frac{1-\delta \tilde{\mu}(\tau ,\rho )}{%
1+\delta \tilde{\mu}(\tau ,\rho )}\right] ^{2}dt^{2}-\frac{[1+\delta \tilde{%
\mu}(\tau ,\rho )]^{4}}{(1+\tfrac{1}{4}\Omega _{k}^{0}\rho ^{2})^{2}}\tilde{a%
}^{2}(\tau )\left[ \mathrm{d}\rho ^{2}+\rho ^{2}\{\mathrm{d}\theta ^{2}+\sin ^{2}\theta \mathrm{d}\phi
^{2}\}\right]
\right\} ,  \notag
\end{eqnarray}%
where $\tilde{\mu}(\tau ,\rho )=\tfrac{1}{4}\left( 2m/R_{s}\right)
^{3/2}(\rho \tilde{a}(\tau ))^{-1}(1+\tfrac{1}{4}\Omega _{k}\rho ^{2})$, $%
\tilde{a}(\tau ):=a(t)$ and $\Omega _{k}^{0}=kH_{0}^{-2}$.

\subsubsection{Interior Metric}

In the interior it is most convenient to work with

\begin{equation*}
R=\left[ 1+\mu \right] (1+\tfrac{1}{4}kr^{2})^{-1/2}a(t)r
\end{equation*}
as the radial coordinate. If $k=0$, then the surface $(t,R)=\mathrm{const}$
has area $4\pi R^{2}$. When $t=t_{0}$ the interior geometry will be
approximately Schwarzschild over scales where $\mathrm{d}R\sim \mathcal{O}%
(R_{s})$ and $\mathrm{d}(t-t_{0})\sim \mathcal{O}(L_{I})$. We define new
dimensionless coordinates by

\begin{equation*}
T=(t-t_{0})/L_{I}\text{ and }\xi =R/R_{s}.
\end{equation*}

In these coordinates the interior metric is:
\begin{eqnarray}
\mathrm{d}s_{int}^{2} &\sim &R_{s}^{2}\left\{ j_{ab}^{(0)}(\xi )+\delta
j_{ab}^{(1)}(\delta T,\xi )+\mathcal{O}\left( \delta ^{2}\right) \right\}
\mathrm{d}x^{a}\mathrm{d}x^{b} \\
&=&R_{s}^{2}\left\{ A(\xi )\left( \frac{R_{s}}{2m}\right) dT^{2}-\left( 1-%
\frac{\delta ^{2}\Omega _{k}(\delta T)}{4}\frac{2m}{R_{s}}X^{2}(\xi )\right)
^{-1}A(\xi )^{-1}\left[ \psi ^{2}+\xi ^{2}\{\mathrm{d}\theta ^{2}+\sin ^{2}\theta \mathrm{d}\phi
^{2}\}\right]
\right\} ,  \notag
\end{eqnarray}%
where $A(\xi )=1-2m/(R_{s}\xi )$,
\begin{equation*}
\psi =\mathrm{d}\xi -\delta h(\delta T)\xi \left( \frac{R_{s}}{2m}\right)
^{1/2}A(\xi )^{1/2}\mathrm{d}T,
\end{equation*}%
and $h(\delta T)=\tilde{a}^{\prime }(\tau =\delta T)/\tilde{a}(\tau )$, $%
\Omega _{k}(\delta T)=\Omega _{k}^{0}\tilde{a}^{-2}(\tau )$. We have defined
$X(\xi )=\frac{\xi }{4}\left( 1+A(\xi )^{1/2}\right) ^{2}$. Note that the
interior metric functions $j_{ab}^{0}$, $j_{ab}^{1}$ etc. can, at each
order, be written in quasi-static form, as functions only of $\xi $ and $%
\delta T$.

\subsubsection{Energy-Momentum Tensor}

The dilaton field theories we are considering couple to the matter sector
through the trace of the energy momentum tensor, $\kappa T_{\mu }^{\mu
}=\kappa (\epsilon -3P)$. In the exterior expansion:
\begin{eqnarray}
H_{0}^{-2}\kappa T_{\mu }^{\mu } &=&\varepsilon _{0}^{e}(\tau )+\delta
\varepsilon _{1}^{e}(\tau ,\rho )+\mathcal{O}(\delta ^{2}) \\
&=&12h^{2}(\tau )+6h^{\prime }(\tau )\left( \frac{1+\delta \tilde{\mu}(\tau
,\rho )}{1-\delta \tilde{\mu}(\tau ,\rho )}\right) +3\Omega _{k}\left( \tau
\right) \left( \frac{2-\delta \tilde{\mu}(\tau ,\rho )}{1-\delta \tilde{\mu}%
(\tau ,\rho )}\right) (1+\delta \tilde{\mu}(\tau ,\rho ))^{-5}-4\Omega
_{\Lambda }^{0}  \notag
\end{eqnarray}%
In the interior expansion we have:
\begin{eqnarray}
R_{s}^{2}\kappa T_{\mu }^{\mu } &=&\delta ^{2}\varepsilon _{2}^{i}(\delta
T,\xi ) \\
&=&12\delta ^{2}\left( \frac{2m}{R_{s}}\right) h^{2}(\delta T)+6\delta
^{2}\left( \frac{2m}{R_{s}}\right) h^{\prime }(\delta T)A(\xi )^{-1/2}
\notag \\
&+&6\delta ^{2}\left( \frac{2m}{R_{s}}\right) \Omega _{k}\left( \delta
T\right) \left( \frac{3+A^{-1/2}(\xi )}{4}\right) \left( \frac{1+A^{1/2}(\xi
)}{2}\right) ^{5}-\delta ^{2}4\left( \frac{2m}{R_{s}}\right) \Omega
_{\Lambda }^{0}  \notag
\end{eqnarray}%
When rewritten in quasi-static form, the only term in the interior expansion
of $T$ is at $\mathcal{O}(\delta ^{2})$.

\subsection{Case II: The flat Gautreau-Tolman-Bondi background}

Gautreau considered the metric outside a massive body embedded in a universe
containing inhomogeneous, spherically symmetric dust and cosmological
constant, $\Lambda $. Unlike the McVittie solution, the dust is in this case
pressureless everywhere. The Gautreau solution is the flat limit of the
Tolman-Bondi model written in curvature coordinates. It is also the $\kappa
P=-\Lambda $ limit of the Leibovitz \cite{Leibovitz}, solutions. Although
not given here, we expect a treatment of dilaton evolution in asymptotically
FRW, locally Schwarzschild, Leibovitz backgrounds to proceed along similar
lines to this case. In co-moving coordinates, which are the most appropriate
for considering the exterior expansion, the metric can be written as
\begin{equation}
\mathrm{d}s^{2}=\mathrm{d}t^{2}-R_{,r}^{2}(t,r)\mathrm{d}r^{2}-R^{2}(t,r)\{%
\mathrm{d}\theta ^{2}+\sin ^{2}\theta \mathrm{d}\phi ^{2}\},  \label{gmetric}
\end{equation}%
where
\begin{equation}
R_{,t}^{2}=\frac{2m+2Z\left( r\right) }{R}+\frac{1}{3}\Lambda R^{2}.
\label{ltbeq.1}
\end{equation}%
The energy density of the dust is given by:
\begin{equation}
\kappa \epsilon =2Z_{,r}/(R^{2}R_{,r}).  \label{lteng}
\end{equation}%
We use the remaining freedom we have in the definition of $R$ to prescribe
that at some epoch of interest, $t=t_{0}$, $r$ is the physical radial
coordinate, so $R(t_{0},r)=r$ and the surface $(t=t_{0},r=const)$ has area $%
4\pi r^{2}$. The surface of our spherical massive body is at $R=R_{s}$. The
quantity $m+Z(r)$ is then defined to be the active gravitational mass
contained inside the shell $(t=t_{0},r)=\mathrm{const}$. and $m$ is the
active gravitational mass of the central massive body at $t=t_{0}$. In
general, the mass of the central object will change over time due to
accretion of external material. We require that as $t\rightarrow \infty $
the central mass remains strictly positive. The dust density must also be
everywhere positive and tend to homogeneity for large $r$. Therefore we have
the conditions:
\begin{eqnarray}
Z(R_{s}) &=&0 \\
Z_{,r} &\geqslant &0 \\
\lim_{r\rightarrow \infty }Z(r) &=&\frac{1}{2}\Omega
_{dust}^{(0)}H_{0}^{2}r^{3} \\
\lim_{r\rightarrow -\infty }Z(r) &>&-m
\end{eqnarray}%
The last requirement is only strictly necessary if we wish the interior to
have the required properties for $t\gg t_{0}$. All we actually require for
this analysis to hold is that the interior remains Schwarzschild\ to leading
order over the time-scales appropriate to our intermediate matching region.
Under our assumptions, equation (\ref{ltbeq.1}) has exact solution:
\begin{eqnarray}
R(t,r) &=&\left( \frac{6(m+Z)}{\Lambda }\sinh ^{2}\left( \frac{\sqrt{%
3\Lambda }}{2}(t-t_{1}(r))\right) \right) ^{1/3},  \label{req.1} \\
t_{1}(r) &=&t_{0}-\frac{2}{\sqrt{3\Lambda }}\sinh ^{-1}\left( \sqrt{\frac{%
r^{3}\Lambda }{6(m+Z)}}\right) .
\end{eqnarray}

\subsubsection{\label{gatextsec} Exterior Metric}

As in the McVittie case, we define dimensionless coordinates $\rho =H_{0}r$
and $\tau =H_{0}t$. With respect to to these coordinates we can then
asymptotically expand $Z(r),$ order by order, in the small parameter $\delta
$. First, we write:
\begin{equation*}
H_{0}Z(r)\sim \frac{1}{2}\Omega _{dust}^{(0)}\rho ^{3}+\delta ^{p}z_{1}(\rho
)+o\left( \delta ^{p}\right) .
\end{equation*}%
Next, we require $2\delta ^{p}z_{1}(\rho )/\Omega _{dust}^{(0)}\rho ^{3}\ll
1 $ for $\rho \sim \mathcal{O}(1)$ so that this is valid asymptotic
expansion; this ensures $p>0$. The unperturbed spacetime is then FRW. Within
the framework of a given model it will be possible to find the value of $p$.
Given the expansion of $Z$ we can then expand $R$ order-by-order using
equation (\ref{req.1}). Putting this expansion back into the metric, eq. (%
\ref{gmetric}), we will have the expanded exterior metric in the form:
\begin{equation*}
\mathrm{d}s_{ext}^{2}\sim H_{0}^{-2}\left( g_{ab}^{(0)}(\tau ,\rho )+\delta
^{p}g_{ab}^{(1)}(\tau ,\rho )+o(\delta ^{p})\right) \mathrm{d}x^{a}\mathrm{d}%
x^{b},
\end{equation*}%
where $g_{ab}^{(0)}$ is the FRW metric.

\subsubsection{\label{gatintsec} Interior Metric}

We define $T=(t-t_{0})/L_{I}$, and $\xi =R/R_{s}$ to be our coordinates in
the interior. We express $Z(r)=Z(\xi ,T)$ and expand out in power of the
small parameter $\delta $:
\begin{equation*}
Z(r)/m\sim \delta ^{q}\mu _{1}(\chi )+o\left( \delta ^{q}\right) .
\end{equation*}%
Where, from eq. (\ref{req.1}), $\chi =\left( \xi ^{3/2}-3T/2\right) ^{2/3}$;
$R_{s}\chi =r+\mathcal{O}(\delta ^{q})$. Putting this expansion back into
the metric, eq. (\ref{gmetric}) gives the first two terms in the asymptotic
expansion of the interior metric:
\begin{equation}
\mathrm{d}s_{int}^{2}\sim R_{s}^{2}\left( j_{ab}^{(0)}(\xi )+\delta
^{q}j_{ab}^{(1)}(\xi ,\chi )+o(\delta ^{q})\right) \mathrm{d}x^{a}\mathrm{d}%
x^{b},  \label{intexp}
\end{equation}%
where $j_{ab}^{(0)}$ is the Schwarzschild metric with mass $m/R_{s}$ in
Painlevé-Gullstrand coordinates (with a re-scaling of the time coordinate):
\begin{equation*}
j_{ab}^{(0)}\mathrm{d}x^{a}\mathrm{d}x^{b}=\frac{R_{s}}{2m}\mathrm{d}%
T^{2}-\left( \mathrm{d}\xi -\xi ^{-1/2}\mathrm{d}T\right) ^{2}-\xi ^{2}%
\{\mathrm{d}\theta ^{2}+\sin ^{2}\theta \mathrm{d}\phi
^{2}\},
\end{equation*}%
and $j_{ab}^{(1)}$ is given by:
\begin{equation*}
j_{ab}^{(1)}\mathrm{d}x^{a}\mathrm{d}x^{b}=\frac{\mu _{1}(\chi )}{\xi ^{1/2}}%
\mathrm{d}\xi \mathrm{d}T-\frac{\mu _{1}(\chi )}{\xi }\mathrm{d}T^{2}.
\end{equation*}%
Unlike in the McVittie solution, the next-to-leading order term in the
interior metric here is not quasi-static, however by defining $\hat{T}=\sqrt{%
R_{s}/2m}T$ to be the time coordinate, the metric can be seen to be
quasi-static in a different sense. Instead of only depending on $T$ through $%
\delta T$, it depends on $\hat{T}$ only through the combination $2m\hat{T}%
/R_{s},$ and for most bodies, with the exception of black holes and neutron
stars, $2m/R_{s}\ll 1$. For eq. (\ref{intexp}) to be a valid asymptotic
expansion we require $\delta ^{q}\mu _{1}(\chi )\ll 1$ for $\chi \sim
\mathcal{O}(1)$.

\subsubsection{Energy - Momentum Tensor}

The Gautreau model is a subcase of the Tolman-Bondi dust models and so the
pressure vanishes, $P=0$, and $T_{\mu }^{\mu }=\epsilon $. In the interior
it is given by:
\begin{eqnarray}
R_{s}^{2}\kappa T_{\mu }^{\mu } &=&\delta ^{q}\varepsilon _{1}^{i}(\xi ,\chi
)+o\left( \delta ^{q}\right) \\
&=&\left( \frac{2m\delta ^{q}}{R_{s}}\right) \frac{\mu _{1}\left( \chi
\right) _{,\chi }}{\xi ^{3/2}\chi ^{1/2}}  \notag
\end{eqnarray}%
An expression for the exterior expansion of $T^{\mu }{}_{\mu }$ is not given
here, but it can be found, with reference to the exact solutions for $R(r,t)$%
, in the same way as the exterior expansion for the metric.

\subsection{Case III: Tolman-Bondi models with simultaneous big bang}

The final class of specific model that we will consider in this paper is,
like the Gautreau solution, also a subcase of the Tolman-Bondi models \cite%
{kr}. This class is distinct from the Gautreau models in that we demand the
big-bang to be simultaneous for all observers, and we do not require the
hyper-surfaces of constant $t$ to be flat. In the limit of spatial
homogeneity these models can therefore reproduce \textit{all} dust-plus-$%
\Lambda $ FRW models. In the Gautreau model $R_{,t}>0$ everywhere. This
implies that the world lines of the matter particles will stream out of the
surface $R=R_{s}$; swept along by the cosmic expansion. In contrast, in this
model we can, and will, require $R_{,t}<0$ near $R=R_{s}$. As a result, dust
particles will fall onto our massive body; this seems a more physical
reasonable scenario than that of the Gautreau case. In co-moving coordinates
the metric can be written as
\begin{equation*}
\mathrm{d}s^{2}=\mathrm{d}t^{2}-\frac{R_{,r}^{2}(t,r)}{1-k(r)}\mathrm{d}%
r^{2}-R^{2}(t,r)\{\mathrm{d}\theta ^{2}+\sin ^{2}\theta \mathrm{d}\phi
^{2}\},
\end{equation*}%
where
\begin{equation*}
R_{,t}^{2}=-k(r)+\frac{2m+2Z\left( r\right) }{R}+\frac{1}{3}\Lambda R^{2}.
\end{equation*}%
The matter content of these models is pressureless dust with a cosmological
constant. The dust energy density is, as in the Gautreau models, given by
equation (\ref{ltbeq.1}). As in the previous example we use the freedom we
have in the definition of $R$ to prescribe that at some epoch of interest, $%
t=t_{0}$, $R$ is the physical radial coordinate i.e. $R(t_{0},r)=r$ and the
surface $(t,R)=const$ always has area $4\pi R^{2}$; this requirement,
combined with the simultaneity of the initial big-bang curvature singularity
, determines the form of $k(r)$. The surface of our massive spherical body
is at $R=R_{s}$. As before, $m+Z(r)$ is the active gravitational mass
interior to the surface $(t=t_{0},r)=\mathrm{const}$ and $m$ is the active
gravitational mass of the central body at $t=t_{0}$. The mass of the central
object will grow over time as a result of accretion. We also require the
dust density to be everywhere positive and tend to spatial homogeneity for
large $r$, hence we need
\begin{eqnarray}
Z(R_{s}) &=&0, \\
Z_{,r} &\geqslant &0, \\
\lim_{r\rightarrow \infty }Z(r) &=&\frac{1}{2}\Omega
_{dust}^{(0)}H_{0}^{2}r^{3}.
\end{eqnarray}%
We want the zeroth-order, exterior limit to be a FRW spacetime with
curvature parameter $k$; this requires $\lim_{r\rightarrow \infty
}k(r)=kr^{2}$. We must require $k(r)>0$ in the interior region; however we
do \emph{not} require $k(r)>0$ everywhere. The exact solution for $R(r,t)$
was found by Barrow and Stein-Schabes and is given in ref. \cite{barrowst}.

\subsubsection{Exterior Metric}

As in the previous examples, we define dimensionless coordinates $\rho
=H_{0}r$ and $\tau =H_{0}t$, and expand $Z(r)$ order by order in the small
parameter $\delta $. We write:
\begin{equation*}
H_{0}Z(r)\sim \frac{1}{2}\Omega _{dust}^{(0)}\rho ^{3}+\delta ^{p}z_{1}(\rho
)+o\left( \delta ^{p}\right) .
\end{equation*}%
We require $2\delta ^{p}z_{1}(\rho )/\Omega _{dust}^{(0)}\rho ^{3}\ll 1$ for
$\rho \sim \mathcal{O}(1)$ to ensure that this is valid asymptotic
expansion; i.e. $p>0$. The unperturbed spacetime is then FRW. The value of $%
p $ is model dependent. With $Z(r)$ specified, we can use the exact
solutions of Barrow and Stein-Schabes, \cite{barrowst, barrowindia}, for $%
R(t,r)$, and the requirement that $R(r,t_{0})=r$ to find the expansion of $%
k(r)$ and from there the expansion of the metric; schematically we have:
\begin{eqnarray}
k(r) &\sim &\Omega _{k}^{(0)}\rho ^{2}+\delta ^{p}E_{1}(\rho )+o\left(
\delta ^{p}\right) , \\
\mathrm{d}s_{ext}^{2} &\sim &H_{0}^{-2}\left( g_{ab}^{(0)}(\tau ,\rho
)+\delta ^{p}g_{ab}^{(1)}(\tau ,\rho )+o(\delta ^{p})\right) \mathrm{d}x^{a}%
\mathrm{d}x^{b},
\end{eqnarray}%
where $g_{ab}^{(0)}$ is the FRW metric.

\subsubsection{\label{ltbsec} Interior Metric}

We take $T=(t-t_{0})/L_{I}$, and $\xi =R/R_{s}$ to be our coordinates in the
interior, and:
\begin{equation*}
Z(r)/m\sim \delta ^{q}\mu _{1}(\eta )+o\left( \delta ^{q}\right) ,
\end{equation*}%
where $\eta =\left( \xi ^{3/2}+3T/2\right) ^{2/3}$; $R_{s}\eta =r+\mathcal{O}%
(\delta ^{q},\delta ^{2/3})$. From the exact solutions we find:
\begin{equation*}
k(\eta )=\delta ^{2/3}k_{0}\left( 1+\delta ^{q}\mu _{1}(\eta )+o\left(
\delta ^{q}\right) \right) +\mathcal{O}\left( \delta ^{5/3}\right) ,
\end{equation*}%
where

\begin{equation*}
k_{0}(\delta T)=\frac{2m}{R_{s}}\left( \frac{\pi }{H_{0}t_{0}+\delta T}%
\right) ^{2/3}.
\end{equation*}%
However, by a redefinition of the $T$ coordinate we can transform away this $%
\mathcal{O}\left( \delta ^{2/3}\right) $ term. We take $T\rightarrow T^{\ast
}$ where:
\begin{equation*}
\sqrt{1-\delta ^{2/3}k_{0}}T^{\ast }=T+\int^{\xi }\frac{\sqrt{\frac{2m}{%
R_{s}\xi ^{\prime }}}\left( 1-\sqrt{1-\left( \frac{\delta ^{2/3}\pi \xi
^{\prime }}{H_{0}t_{0}+\delta T}\right) }\right) }{1-\frac{2m}{R_{s}\xi
^{\prime }}}\mathrm{d}\xi ^{\prime }.
\end{equation*}%
To leading order, we find $T\sim T^{\ast }$. The interior expansion of the
metric is written:
\begin{equation*}
\mathrm{d}s_{int}^{2}\sim R_{s}^{2}\left( j_{ab}^{(0)}(\xi )+\delta
^{q}j_{ab}^{(1)}(\xi ,\chi )+o(\delta ^{q})\right) \mathrm{d}x^{a}\mathrm{d}%
x^{b}+o(\delta ^{q}).
\end{equation*}%
where $j_{ab}^{(0)}$ and $j_{ab}^{(1)}$ are given by:
\begin{eqnarray}
j_{ab}^{(0)}\mathrm{d}x^{a}\mathrm{d}x^{b} &=&\frac{R_{s}}{2m}\mathrm{d}%
T^{\ast2}-\left( \mathrm{d}\xi +\xi ^{-1/2}\mathrm{d}T^{\ast}\right) ^{2}-\xi ^{2}%
\{\mathrm{d}\theta ^{2}+\sin ^{2}\theta \mathrm{d}\phi
^{2}\},  \label{j1eq.2} \\
j_{ab}^{(1)}\mathrm{d}x^{a}\mathrm{d}x^{b} &=&-\frac{\mu _{1}(\chi )}{\xi
^{1/2}}\mathrm{d}\xi \mathrm{d}T-\frac{\mu _{1}(\chi )}{\xi }\mathrm{d}T^{2}.
\end{eqnarray}%
As with the Gautreau case, $j_{ab}^{(0)}$ is the Schwarzschild metric in
Painlevé-Gullstrand coordinates. The discussion about the quasi-static
nature of the Gautreau $j_{ab}^{(1)}$ term, given at the end of section \ref%
{gatintsec}, applies equally well here. For the approximation above to be a
valid asymptotic expansion we require $\delta ^{q}\mu _{1}(\chi )\ll 1$ for $%
\chi \sim \mathcal{O}(1)$.

\subsubsection{Energy - Momentum Tensor}

The energy density of the dust locally is given by:
\begin{eqnarray}
R_{s}^{2}\kappa T_{\mu }^{\mu } &=&\delta ^{q}\varepsilon _{1}^{i}(\xi ,\chi
)+o\left( \delta ^{q}\right) \\
&=&\left( \frac{2m\delta ^{q}}{R_{s}}\right) \frac{\mu _{1}\left( \chi
\right) _{,\chi }}{\xi ^{3/2}\chi ^{1/2}}+o\left( \delta ^{q}\right) .
\notag
\end{eqnarray}%
We will not give the exterior expansion of $T_{\mu }^{\mu }$ here
explicitly, although it can in principle be found with reference to the
exact solutions for $R(r,t)$ if required.

\subsection{Boundary Conditions}

As the physical radius tends to infinity, $R\rightarrow \infty $, we demand
that the dilaton tends to its homogeneous cosmological value: $\phi
(R,t)\rightarrow \phi _{c}(t)$. We can apply this boundary condition in the
exterior region but not in the interior. To solve the interior dilaton field
equations we need to specify the dilaton-flux passing out from the surface
of our `star' at $R=R_{s}$. At leading order we parametrise this by:
\begin{equation}
-R_{s}^{2}\left( 1-\frac{2m}{R_{s}}\right) \left. \partial _{R }\phi
_{0}\right\vert _{R =R_{s}}=2mF\left( \bar{\phi}_{0}\right)
=\int_{0}^{R_{s}}\mathrm{d}R^{\prime }R^{\prime }{}^{2}B_{,\phi }(\phi _{0}(%
R^{\prime }))\kappa \epsilon (R^{\prime }),  \label{phiflux}
\end{equation}%
where $\bar{\phi}_{0}=\phi _{0}(R=R_{s})$. The function $F(\phi )$ can be
found by solving the dilaton field equations to leading order in the $%
R<R_{s} $ region. If the interior region is a black-hole ($R_{s}=2m$) then
we must have $F(\phi )=0$; otherwise we expect $F(\phi )\sim B_{,\phi }(\phi
)$. Without considering the sub-leading order dilaton evolution inside our
`star', i.e. at $R<R_{s}$, we cannot rigorously specify any boundary
conditions beyond leading order. Despite this, we can guess at a general
boundary condition by perturbing eq. (\ref{phiflux}):
\begin{equation}
-R_{s}^{2}\left( 1-\frac{2m}{R_{s}}\right) \left. \partial _{R}\tilde{\delta}%
(\phi )\right\vert _{\xi =R_{s}}=-\left. \tilde{\delta}\left( \sqrt{-g}%
g^{RR}\right) \partial _{R}\phi _{0}\right\vert _{R=R_{s}}+2\tilde{\delta}%
(M)F\left( \bar{\phi}_{0}\right) +2mF_{,\phi }(\bar{\phi}_{0})\tilde{\delta}%
\left( \bar{\phi}_{0}\right) +\mathrm{smaller}\;\mathrm{terms},
\label{pertbdry}
\end{equation}%
where $\tilde{\delta}(X)$ is the first sub-leading order term in the
interior expansion of $X$; $M$ is the total mass contained inside $\xi
<R_{s} $ and is found by requiring the conservation of energy; at $t=t_{0}$
we have $M=m$. Only $\tilde{\delta}\left( \bar{\phi}_{0}\right) $ remains
unknown, however we shall assume it to be the same order as $\tilde{\delta}%
(\phi )$ and see that this unknown term is usually suppressed by a factor of
$2m/R_{s} $ relative to the other terms in eq. (\ref{pertbdry}).

\section{Applications}

\subsection{\label{zeroth} Zeroth-order solutions}

\label{matchsec} In accord with our prescription, all of the models that we
have considered share the property that, to lowest order in $\delta $, the
interior is pure Schwarzschild, and the exterior is pure FRW.

\subsubsection{Exterior}

In the exterior, the dilaton field takes its cosmological value: $\phi =\phi
(t)_{c}$ to zeroth order and so satisfies the homogeneous conservation
equation:
\begin{equation*}
\partial _{t}^{2}\phi (t)_{c}+3H\partial _{t}\phi (t)_{c}=B_{,\phi }\left(
\phi _{c}\right) \kappa \epsilon _{dust}^{c}(t)-V_{,\phi }\left( \phi
_{c}\right) ,
\end{equation*}%
The effect of the interior region on the exterior should, even for finite $%
\delta $, become increasingly smaller as $r\rightarrow \infty $. As a result
there will be no sub-leading order, $r$-independent, terms in the exterior
expansion of $\phi $. Equivalently, the homogeneous mode of $\phi $ in the
exterior will be given by the cosmological term, $\phi _{c}$, alone to all
orders

\subsubsection{Interior}

In the interior the dilaton field satisfies the wave equation on a
Schwarzschild background at zeroth order in $\delta $; we take $\phi $ to be
quasi-static to leading order. By applying the zeroth-order boundary
condition, eq. (\ref{phiflux}), we find:
\begin{equation*}
\phi _{0}=\phi _{e}\left( \delta T\right) -F\left( \bar{\phi}_{0}\right) \ln
\left( 1-\frac{2m}{R_{s}\xi }\right) ,
\end{equation*}%
where $\phi _{e}\left( \delta T\right) $ is a constant of integration to be
determined by the matching procedure.

\subsubsection{Matching}

As specified above, we should rewrite the exterior and interior expansion of
$\phi $ in some intermediate region, with length scale, $L_{I}<L_{int}<L_{E}$
. We define

\begin{equation*}
L_{int}\mathcal{T}=(t-t_{0}).
\end{equation*}

The only homogeneous mode in the exterior expansion was the cosmological
term, $\phi _{c}(t)=\phi _{hom}\left( L_{int}\mathcal{T}/L_{E}\right) $.
This term will therefore appear at leading order in the intermediate
expansion. Any other homogeneous terms that result from taking the
intermediate limit of the exterior expansion must be sub-leading order.
Therefore, we conclude that $\phi _{hom}=\phi _{c}(t)$ is the \emph{only}
leading order, homogeneous term in the intermediate expansion of $\phi $. In
addition, all sub-leading-order homogeneous terms must also depend on time only $t$, and so will be quasi-static ($L_{int}\mathcal{T%
}/L_{E}$ dependent) in the intermediate regions.

When the intermediate limit of the interior approximation is taken, it is
clear that homogeneous terms in the interior will map to homogeneous terms
in the intermediate zone. The matching criteria therefore implies that all
homogeneous terms in the interior must be quasi-static, i.e. only depending
on time through $\delta T$, and that at leading order:
\begin{equation*}
\phi _{e}\left( \delta T\right) =\phi _{e}\left( L_{int}\mathcal{T}%
/L_{E}\right) =\phi _{hom}\left( L_{int}\mathcal{T}/L_{E}\right) =\phi
_{c}(t)
\end{equation*}%
The interior solution therefore reads:
\begin{equation*}
\phi _{0}=\phi _{c}\left( t\right) -F\left( \bar{\phi}_{0}\right) \ln \left(
1-\frac{2m}{R_{s}\xi }\right) ,
\end{equation*}%
and $\bar{\phi}_{0}=\phi _{c}\left( t\right) -F\left( \bar{\phi}_{0}\right)
\ln \left( 1-\frac{2m}{R_{s}}\right) $. We see directly the effect of the
cosmological evolution of \ $\phi $ on the local region.

\subsection{Case I: The McVittie Background}

When the matched asymptotic expansion method is applied to the McVittie
background, the analysis goes through in much the same way as it did in
example 2 above. In the interior we find:
\begin{equation*}
\phi _{I}=\phi _{I}^{(0)}+\delta ^{2}\frac{2m}{R_{s}}\phi _{I}^{(1)}+%
\mathcal{O}\left( \delta ^{4}\right) ,
\end{equation*}%
\noindent where $\phi _{I}^{(1)}$ satisfies:
\begin{eqnarray}
\frac{1}{\xi ^{2}}\partial _{\xi }\left( \xi ^{2}A(\xi )\partial _{\xi }\phi
_{I}^{(1)}\right) &=&\frac{\phi _{E}^{^{\prime \prime }}\left( \delta
T\right) }{A(\xi )}+h\left( \delta T\right) \left( \frac{3}{A^{1/2}(\xi )}+%
\frac{1}{2}\frac{2m}{R_{s}RA^{3/2}(\xi )}\right) \phi _{E}^{\prime }\left(
\delta T\right) \\
&&-\frac{2m}{R_{s}}\frac{h^{\prime }(\delta T)}{A^{1/2}(\xi )\xi }F\left(
\bar{\phi}_{0}\right) +\frac{2m}{R_{s}\xi ^{2}}\partial _{\xi }\left( \frac{%
\Omega _{k}(\delta T)X^{2}}{8}-\frac{h^{2}\xi ^{2}}{A(\xi )}\right) F\left(
\bar{\phi}_{0}\right)  \notag \\
&&-B_{,\phi }\left( \phi _{I}^{(0)}\right) \left[ 12h^{2}+6h^{\prime }A(\xi
)^{1/2}+3\Omega _{k}(\delta T)\left( \frac{1+3A^{1/2}}{A^{1/2}}\right)
\left( \frac{1+A^{1/2}}{2}\right) ^{5}-12\Omega _{\Lambda }\right]  \notag \\
&&+R_{s}^{2}V_{,\phi }\left( \phi _{I}^{0}\right) .  \notag
\end{eqnarray}%
This can be integrated once the functions $B_{,\phi }$ and $V_{,\phi }$ are
specified. If we also have
\begin{eqnarray}
\frac{B_{,\phi \phi }(\phi _{E})F(\bar{\phi}_{0})\frac{2m}{R_{s}\xi }}{%
B_{,\phi }(\phi _{E})} &\ll &1, \\
\frac{V_{,\phi \phi }(\phi _{E})F(\bar{\phi}_{0})\frac{2m}{R_{s}\xi }}{%
V_{,\phi }(\phi _{E})} &\ll &1,
\end{eqnarray}%
then we can solve for $\phi _{I}^{(1)}$ as an asymptotic series in $2m/R_{s}$
(provided this quantity is small). These relations will almost always hold
provided that $2m/R_{s}\xi \ll 1$ and the cosmological value of $\phi $ does
not lie near the minimum of $V$ or $B$. To $\mathcal{O}\left(
2m/R_{s}\right) $ we find:
\begin{eqnarray}  \label{mcintexp}
\phi _{I}^{(1)} &\sim &\frac{2m\xi }{R_{s}}\left( \tfrac{1}{2}\left( \phi
_{E}^{(0)\prime \prime }+h\phi _{E}^{(0)\prime }-h^{\prime }F\left( \bar{\phi%
}_{0}\right) \right) +\tfrac{1}{8}\Omega _{k}(\delta T)F\left( \bar{\phi}%
_{0}\right) -h^{2}F\left( \bar{\phi}_{0}\right) \right. \\
&&\left. +\frac{3}{2}B_{,\phi }(\phi _{E})\left( h^{\prime }(\delta T)+%
\tfrac{11}{4}\Omega _{k}(\delta T)\right) -\frac{1}{2}F\left( \bar{\phi}%
_{0}\right) B_{,\phi \phi }\left( \phi _{E}\right) R_{s}^{2}\kappa \epsilon
_{dust}^{(0)}+F\left( \bar{\phi}_{0}\right) R_{s}^{2}V_{,\phi \phi }\left(
\phi _{E}\right) \right)  \notag \\
&&+\frac{2m}{R_{s}}\left( \frac{C(\delta T)}{\xi }+D(\delta T)\right) +%
\mathcal{O}\left( \frac{2m}{R_{s}}\right) ^{2},  \notag
\end{eqnarray}%
where $C(\delta T)$ is determined by the boundary conditions on $R=R_{s}$
and $D(\delta T)$ comes from the matching conditions. The matching procedure
would set $D(\delta T)=0$ if we were to continue our asymptotic expansions to
an order smaller than we do here. We do not expect $C(\delta T)$ to be any
larger than the other terms in this expression when $\xi =1$. The exterior
expansion, to $\mathcal{O}(\delta )$, along with the matching conditions
that determine the otherwise unknown functions in it, is given in Appendix %
\ref{appA}. In this appendix, we show that these matching conditions are
self-consistent, and hence that, to the order considered, the matching
procedure works correctly. As with the simple example we considered in
section \ref{scalexp}, we see that when the matching conditions are applied,
the background sets the form of the homogeneous terms in the interior
solution, and the interior solution gives us the behaviour of the
inhomogeneous terms in the exterior expansion. This is to be expected, since
in these models the interior region is the sole source of the
inhomogeneities in both the spacetime and the dilaton field. This picture
would be more complicated if we were to consider more than one interior
region, as in the case of a nested series of shells of matter of differing
densities.

Our original concern was the behaviour of the time derivative of $\phi $ in
the interior and we have shown that in the McVittie background:
\begin{equation*}
\frac{\partial _{t}\phi _{I}}{\partial _{t}\phi _{c}}\sim 1-F_{,\phi }\left(
\bar{\phi}_{0}\right) \ln (1-2m/R)+\mathcal{O}\left( \delta ^{2}\left(
2m/R)\right) ^{2}\right) .
\end{equation*}%
Since, in most cases of interest, the dilaton to matter coupling is weak and
we are far from the Schwarzschild radius of our `star', both $F(\left( \bar{%
\phi}_{0}\right) $ and $2m/R$ are $\ll 1$. Hence: $\partial _{t}\phi
_{I}/\partial _{t}\phi _{c}\approx 1$, and the local time evolution of the
dilaton tracks the cosmological one. The strength of this result arises
partly from our restrictive choice of background metric. If the background
is taken to be Tolman-Bondi rather than McVittie, we can find quite
different behaviour.\emph{\ }

\subsection{\label{gatsec} Case II: The Gautreau-Tolman-Bondi Background}

We assume that we have specified some $Z(r;\delta )$ that has outer and
inner approximations as given in sections \ref{gatextsec} and \ref{gatintsec}
respectively. But we must be careful to ensure that the form of $Z(r;\delta
) $ is such that these two approximations can be matched in some
intermediate region. We can now proceed to solve the dilaton evolution
equation in the inner and outer limits and then apply our matching
procedure. We have already seen that at zeroth order the matching conditions
imply that $\phi $ is approximated by
\begin{eqnarray}
&&\mathrm{inner}\;\mathrm{approx}\sim \phi _{I}^{(0)}(\xi ;\delta T)+%
\mathcal{O}\left( \delta ^{q}\right) =\phi _{c}\left( \tau =\tau _{0}+\delta
T\right) -F\left( \bar{\phi}_{0}\right) \ln \left( 1-\frac{2m}{R_{s}\xi }%
\right) +\mathcal{O}\left( \delta ^{q}\right) ,  \notag \\
&&\mathrm{outer}\;\mathrm{approx}\sim \phi _{c}\left( \tau \right) +\mathcal{%
O}\left( \delta ^{p}\right) .  \notag
\end{eqnarray}%
We now consider the next-to-leading order terms with a view to determining
how time variation of $\phi $ in the interior is related to its cosmological
rate of change.

In the interior the local energy density and metric are, to $\mathcal{O}%
\left( \delta ^{q}\right) $, depend on $T$ only through $\chi $ and are
written as functions of $\xi $ and $\chi $. The calculation is easiest if we
move from $(T,\xi )$ coordinates to $(T,\chi )$ ones and write:
\begin{equation*}
\phi \sim \phi _{I}^{(0)}(\xi ;\delta T)+\delta ^{q}\phi _{I}^{(1)}(T,\chi
;\delta T)+o\left( \delta ^{q}\right) ,
\end{equation*}%
where $\phi _{I}^{1}$ satisfies (for $q<2$):
\begin{eqnarray}  \label{phiintev}
-\frac{2m}{R_{s}}\left( \xi ^{3/2}\phi _{I,TT}^{(1)}+\frac{3}{2}\phi
_{I,T}^{(1)}\right) &+&\frac{1}{\chi ^{1/2}}\left( \frac{\xi ^{5/2}}{\chi
^{1/2}}\phi _{I,\chi }^{(1)}\right) _{,\chi }=-\frac{2m}{R_{s}}B_{,\phi
}\left( \phi _{I}^{0}\right) \frac{\mu _{1}\left( \chi \right) _{,\chi }}{%
\chi ^{1/2}}  \label{gintphieq.} \\
&+&\left( \frac{2m}{R_{s}}\right) ^{2}F\left( \bar{\phi}_{0}\right) \left[
\frac{\mu _{1}\left( \chi \right) _{,\chi }}{\xi \chi ^{1/2}\left( 1-\tfrac{%
2m}{R_{s}\xi }\right) }-\frac{2\mu _{1}\left( \chi \right) }{\xi
^{5/2}\left( 1-\tfrac{2m}{R_{s}\xi }\right) ^{2}}\right] .
\end{eqnarray}%
There is also a single term that appears at order $\delta $ which we have
omitted from the above expression; we shall deal with this later. In
general, the above equation is difficult to solve exactly, however, in most
of the cases of interest the surface of our `star' will be far outside its
Schwarzschild radius and so $2m/R_{s}\ll 1$. We may then solve eq. (\ref%
{gintphieq.}) as an asymptotic series in $2m/R_{s}$, valid where $\xi \gg
\frac{2m}{R_{s}}$. Given the nature of our problem we are only interested in
the leading term:
\begin{equation*}
\phi _{I}^{(1)}\sim -\frac{2m}{R_{s}}B_{,\phi }\left( \phi _{c}\right)
\left( \int^{\chi }\mathrm{d}\chi ^{\prime }\frac{\mu _{1}\left( \chi
^{\prime }\right) _{,\chi }}{\xi (\chi ^{\prime },T)}-\frac{\mu _{1}\left(
\chi \right) }{\xi }+\frac{A(T)}{\xi }+C(T)\right) +\mathcal{O}\left( \left(
\frac{2m}{R_{s}}\right) ^{2}\right) .
\end{equation*}%
Here, $A(T)$ and $B(T)$ are `constants' of integration; $A(T)$ should be
determined by a boundary condition on the surface at $R=R_{s}$ i.e. at $\xi
=1$, and $B(T)$ will come out of the matching. We argued in section \ref%
{matchsec} that the matching requires that all homogeneous terms in the
interior expansion be quasi-static, thus $B(T)=\hat{B}(\delta T)$. Without
knowing more about the region $\xi <1$, we do not have a boundary condition
capable of determining $A(T)$. If, however, we assume that the prescription
given by eq. (\ref{pertbdry}) is at least approximately correct then we can
proceed. We assume that the next-to-leading order perturbation of $\bar{\phi}%
_{0}$ occurs at the same order in $\delta $ and $2m/R_{s}$ as the
next-to-leading order term in $\phi $. Given these assumptions we prescribe
our full boundary condition at $R=R_{s}$ to be given by eq. (\ref{phiflux})
but with $\bar{\phi}_{0}\rightarrow \bar{\phi}_{0}+\mathcal{O}\left( \delta
^{q}2m/R_{s}\right) $ and $m\rightarrow m+Z|_{R=R_{s}}$. The resultant
boundary condition on $\phi _{I}^{(1)}$ is then remarkably simple:
\begin{equation*}
\partial _{\xi }\phi _{I}^{(1)}|_{\xi =1}=-\frac{2m}{R_{s}}F(\bar{\phi}%
_{0})\mu _{1}\left( \chi (\xi =1,T)\right) .
\end{equation*}%
With this choice of boundary condition,

\begin{equation*}
A(T)=\left( 1-F(\bar{\phi}_{0})/B(\phi _{c})\right) \mu _{1}\left( \chi (\xi
=1,T\right)
\end{equation*}
and the interior solution is:
\begin{equation*}
\phi _{I}^{(1)}\sim -\frac{2m}{R_{s}}B_{,\phi }\left( \phi _{c}\right)
\left( \int^{\chi }\mathrm{d}\chi ^{\prime }\frac{\mu _{1}\left( \chi
^{\prime }\right) _{,\chi }}{\xi (\chi ^{\prime },T)}-\frac{\mu _{1}\left(
\chi \right) }{\xi }+\left( 1-\frac{F(\bar{\phi}_{0})}{B(\phi _{c})}\right)
\frac{\mu _{1}\left( \chi (\xi =1,T\right) }{\xi }+\hat{B}(\delta T)\right) +%
\mathcal{O}\left( \left( \frac{2m}{R_{s}}\right) ^{2}\right) .
\end{equation*}%
The unknown quasi-static term, $\hat{B}\left( \delta T\right) $, will not
affect the leading-order time-variation of $\phi _{I}^{(1)}$ and so will not
have any bearing on our result. We could, in principle, solve for $\phi $ in
the outer approximation, however, since we are interested only in dynamics
of the dilaton field in the interior, but we will not do so.

Recall that we dropped an order-$\delta $ term in writing down eq. (\ref%
{phiintev}); this perturbation decouples from the one at order $\delta ^{p}$%
, and is quasi-static. Writing

\begin{equation*}
\phi _{I}\sim \phi _{I}^{(0)}(\delta T)+\delta ^{p}\phi _{I}^{(1)}+\delta
\phi _{I}^{(\delta )}(\xi ;\delta T)
\end{equation*}
we find:
\begin{equation*}
\frac{1}{\xi ^{2}}\left( \xi (\xi -2m/R_{s})\phi _{I,\xi }^{(\delta
)}\right) _{,\xi }=\frac{2m}{R_{s}}\frac{1}{\xi ^{2}}\left( -\sqrt{\frac{1}{%
\xi }}\xi ^{2}\right) _{,\xi }\left( \phi _{e}^{\prime }\left( \delta
T\right) -F_{,\phi }\left( \bar{\phi}_{0}\right) \ln \left( 1-\frac{2m}{%
R_{s}\xi }\right) \bar{\phi}_{0}^{\prime }(\delta T)\right) .
\end{equation*}%
Solving this we find:
\begin{eqnarray}
\phi _{I}^{(\delta )} &=&2\left( \frac{2m}{R_{s}}\right) ^{3/2}\left( \sqrt{%
\frac{R_{s}\xi }{2m}}+\frac{1}{2}\ln \left\vert \frac{\sqrt{\frac{R_{s}\xi }{%
2m}}-1}{\sqrt{\frac{R_{s}\xi }{2m}}+1}\right\vert \right) \phi _{e}^{\prime
}\left( \delta T\right) -\left( \frac{2m}{R_{s}}\right) ^{3/2}l(\xi
)F_{,\phi }\left( \bar{\phi}_{0}\right) \bar{\phi}_{0}^{\prime }\left(
\delta T)\right)  \label{phidel} \\
&&-\left( \frac{2m}{R_{s}}\right) ^{3/2}A\ln \left( 1-\frac{2m}{R_{s}\xi }%
\right) .  \notag
\end{eqnarray}%
\noindent where $A$ is a constant of integration and
\begin{equation*}
l(\xi )=-\int^{\xi }\mathrm{d}\xi ^{\prime }\xi ^{\prime }{}^{-2}\left( 1-%
\frac{2m}{R_{s}\xi ^{\prime }}\right) ^{-1}\left( \left( \frac{R_{s}\xi }{2m}%
\right) ^{3/2}\ln \left( 1-\frac{2m}{R_{s}\xi ^{\prime }}\right) -2\left(
\frac{R_{s}\xi }{2m}\right) ^{1/2}-2\ln \left\vert \frac{1+\sqrt{1-\frac{2m}{%
R_{s}\xi ^{\prime }}}-\sqrt{\frac{2m}{R_{s}\xi ^{\prime }}}}{1+\sqrt{1-\frac{%
2m}{R_{s}\xi ^{\prime }}}+\sqrt{\frac{2m}{R_{s}\xi ^{\prime }}}}\right\vert
\right) .
\end{equation*}%
The term proportional to $l(\xi )$ dies off as $\xi ^{1/2}$, and is
suppressed by a factor of $2m/R_{s}\xi $ relative to the first term in eq. (%
\ref{phidel}). Both these terms only depend on time through $\delta T$ and
are thus deemed quasi-static. These terms will \emph{not}, therefore, change
the leading-order behaviour of the time derivative of $\phi $ in the
interior.

Although we are concerned mostly with objects which are much larger than
their Schwarzschild radii it would be nice to be able to address the problem
of black-hole gravitational memory via this method. For the zeroth-order
approximation to $\phi $ to be well-defined on the horizon we need $F(\bar{%
\phi}_{0})=0$ (this is just a statement of the `no-hair' theorem for
Schwarzschild black holes). We can remove any divergence in $\phi
_{I}^{(\delta )}$ near the horizon by an appropriate choice of the constant
of integration, $A$. We take $A=\phi _{e}^{\prime }(\delta T)$. This is
analogous to what was done by Jacobson in \cite{jacobson}. We can now absorb
the first and last terms on the RHS of eq. \ref{phidel} into the definition
of $\phi _{e}(\delta T)$ by a definition of the time coordinate:
\begin{eqnarray*}
&&\phi _{e}(\delta T)\rightarrow \phi _{e}(\delta \tilde{T}) \\
\mathrm{where}\; &&\tilde{T}=T+2\left( \frac{2m}{R_{s}}\right) ^{3/2}\left(
\sqrt{\frac{R_{s}\xi }{2m}}-\ln \left\vert 1+\sqrt{\frac{2m}{R_{s}\xi }}%
\right\vert \right) .
\end{eqnarray*}%
The zeroth-order matching is not affected since $\delta (\tilde{T}-T)$ will
remain sub-leading order in any intermediate region. We note that $L_{I}\hat{%
T}=v-R-2m\ln (R/2m)$ where $v=t_{s}+R+2m\ln (R/2m-1)$ and $t_{s}$ is the
usual Schwarzschild time coordinate. Thus the leading-order homogeneous term
in $\phi _{I}$ has the same behaviour as that predicted by Jacobson (see
ref. \cite{jacobson} and section \ref{jacsec} above).

By expanding $\phi _{e}$ w.r.t. the co-moving time, $L_{I}T$, rather than the
Schwarzschild time, $t_{s}$, as Jacobson did, we avoid one of the problems
encountered in his analysis: that the time coordinate use to make the
expansion diverges on the horizon.

At lower orders we can always ensure, by choice of the constants of
integration, that $\lim_{R\rightarrow 0}(1-2m/R)\partial _{R}\phi |_{T}=0$.
This boundary condition ensures that the inner approximation does not
diverge as $R\rightarrow 2m$, and remains valid on the horizon. This
justifies the statement that, if $p>1$, Jacobson's prediction is itself a
valid asymptotic approximation to $\phi $ on the horizon. The condition $p>1$
is equivalent to: $\epsilon _{local}/\epsilon _{c}\ll 1/(2mH_{0})(\gg 1)$,
with $\epsilon _{l}$ being the average value of the energy density, outside
the black-hole, over length scales $\mathcal{O}(2m)$ from the horizon.

We are now in a position to study the conditions under which the local time
variation of $\phi $ tracks its cosmological value. In the interior we have:
\begin{eqnarray}
\phi _{I,\tilde{T}} &\approx &\delta \phi _{e}^{\prime }(\delta \tilde{T}%
)\left( 1-F_{,\phi }\left( \bar{\phi}_{0}\right) \ln \left( 1-\frac{2m}{%
R_{s}\xi }\right) \right)  \label{gatphitev} \\
&&+\delta ^{q}B_{,\phi }(\phi _{c})\frac{2m}{R_{s}}\left[ \int^{\chi }%
\mathrm{d}\chi ^{\prime }\frac{\mu _{1}\left( \chi ^{\prime }\right) _{,\chi
}}{\xi ^{5/2}(\chi ^{\prime },T)}+\left( 1-\frac{F(\bar{\phi}_{0})}{B(\phi
_{c})}\right) \frac{\mu _{1,\chi }(\xi =1,T)}{\xi ^{3/2}}\right] +...  \notag
\end{eqnarray}%
\emph{The excluded terms are then certainly smaller than the second term,
however, since they might be larger than the first term, both numerically
and in the limit }$\delta \rightarrow 0$\emph{\ , the above expression is
not a formal asymptotic approximation.} When the first term in eq. (\ref%
{gatphitev}) dominates, condition (\ref{wettcond}) holds, and the local time
evolution of the scalar field is, to leading order, the same as the
cosmological one. Assuming our background choice is suitable for the
application of this method, condition (\ref{wettcond}) fails to hold if, and
only if, the second term in this expression dominates over the first.

\subsection{Case III: Tolman-Bondi models with simultaneous big bang}

In the interior, the leading-order corrections to the metric occur at either
order $\delta ^{q}$ or $\delta ^{2/3}$. The order $\delta ^{2/3}$ correction
comes from the leading, quasi-static, and $\xi $-independent term in the
expansion of $k(r)$; we have seen that this can be transformed away by a
redefinition of the local time coordinate, $T\rightarrow T^{\ast }$, as done
in section \ref{ltbsec}.

In the intermediate region $T\sim \sqrt{1-\delta ^{2/3}k_{0}}T^{\ast }$, the
zeroth-order matching goes through in the same way as it did it the two
previous cases. The zeroth-order interior solution is%
\begin{equation*}
\phi _{0}=\phi _{e}\left( \delta \sqrt{1-\delta ^{2/3}k_{0}}T^{\ast }\right)
-F\left( \bar{\phi}_{0}\right) \ln \left( 1-\frac{2m}{R_{s}\xi }\right) .
\end{equation*}%
The analysis of the $\mathcal{O}\left( \delta ^{q}\right) $ terms then
follows through in much the same way as for the Gautreau case, but with
\begin{equation*}
\chi =\left( \xi ^{3/2}-3T/2\right) ^{2/3}\rightarrow \eta =\left( \xi
^{3/2}+3T/2\right) ^{2/3}.
\end{equation*}

One finds that the $\mathcal{O}\left( \delta ^{q}\right) $ correction to the
interior solution is given by
\begin{equation*}
\phi _{I}^{(1)}\sim -\frac{2m}{R_{s}}B_{,\phi }\left( \phi _{c}\right)
\left( \int^{\eta }\mathrm{d}\eta ^{\prime }\frac{\mu _{1}\left( \eta
^{\prime }\right) _{,\chi }}{\xi (\eta ^{\prime },T)}-\frac{\mu _{1}\left(
\eta \right) }{\xi }+\left( 1-\frac{F(\bar{\phi}_{0})}{B(\phi _{c})}\right)
\frac{\mu _{1}\left( \eta (\xi =1,T\right) }{\xi }+\hat{B}(\delta T)\right) +%
\mathcal{O}\left( \left( \frac{2m}{R_{s}}\right) ^{2}\right) .
\end{equation*}%
There is also a quasi-static order $\delta $ correction to $\phi _{I}$, just
as for the Gautreau case. We find
\begin{eqnarray*}
\phi _{I}^{(\delta )} &=&-\sqrt{1-\delta ^{2/3}k_{0}}2\left( \frac{2m}{R_{s}}%
\right) ^{3/2}\left( \sqrt{\frac{R_{s}\xi }{2m}}+\frac{1}{2}\ln \left\vert
\frac{\sqrt{\frac{R_{s}\xi }{2m}}-1}{\sqrt{\frac{R_{s}\xi }{2m}}+1}%
\right\vert \right) \phi _{e}^{\prime }\left( \delta T^{\ast} \right) +\left(
\frac{2m}{R_{s}}\right) ^{3/2}l(\xi )F_{,\phi }\left( \bar{\phi}_{0}\right)
\bar{\phi}_{0}^{\prime }\left( \delta T)\right) \\
&&+\left( \frac{2m}{R_{s}}\right) ^{3/2}A\ln \left( 1-\frac{2m}{R_{s}\xi }%
\right) .
\end{eqnarray*}%
\noindent with $l(\xi )$ as given in section \ref{gatsec}. As above, we
shall choose the constant of integration, $A$, so as to make $\phi
_{I}^{\delta }$ well-defined as $R_{s}\xi \rightarrow 2m$ whenever $F\left(
\bar{\phi}_{0}\right) =0$; we can then absorb this correction into the
definition of $\phi _{e}$:
\begin{equation*}
\phi _{e}\left( \delta \sqrt{1-\delta ^{2/3}k_{0}}T^{\ast }\right)
\rightarrow \phi _{e}\left( \delta \sqrt{1-\delta ^{2/3}k_{0}}\tilde{T}%
^{\ast }\right) .
\end{equation*}%
The transform $T^{\ast }\rightarrow \tilde{T}^{\ast }$ is the same as the
one for $T\rightarrow \tilde{T}$ given in section \ref{gatsec}; $L_{I}\tilde{%
T}=v-R-2m\ln (R/2m)$. As in previous case, we can now analyse the conditions
under which the local time variation of $\phi $ tracks its cosmological
value:
\begin{eqnarray}
\phi _{I,\tilde{T}\ast } &\approx &\delta \sqrt{1-\delta ^{2/3}k_{0}}\phi
_{e}^{\prime }(\sqrt{1-\delta ^{2/3}k_{0}}\delta \tilde{T}^{\ast} )\left(
1-F_{,\phi }\left( \bar{\phi}_{0}\right) \ln \left( 1-\frac{2m}{R_{s}\xi }%
\right) \right)  \label{ltbphitev} \\
&&-\delta ^{q}B_{,\phi }(\phi _{c})\frac{2m}{R_{s}}\left[ \int^{\eta }%
\mathrm{d}\eta ^{\prime }\frac{\mu _{1}\left( \eta ^{\prime }\right) _{,\eta
}}{\xi ^{5/2}(\eta ^{\prime },T)}+\left( 1-\frac{F(\bar{\phi}_{0})}{B(\phi
_{c})}\right) \frac{\mu _{1,\eta }(\xi =1,T^{\ast} )}{\xi ^{3/2}}\right] +...
\notag
\end{eqnarray}%
This not a formal asymptotic approximation, but the excluded terms are
certainly smaller than at least one of the two terms in the above expansion.
The effect of the transform $T\rightarrow T^{\ast }$ is to induce a slight,
sub-leading order, lag in the time evolution of the scalar field, that
increases as $\xi $ decreases. As in the Gautreau case above, condition (\ref%
{wettcond}) will fail to hold if, and only if, the second term in this
expansion dominates. We will interpret this condition, and the similar one
which arose from the analysis of the Gautreau case, in terms of what it
requires for the local energy-density in section \ref{intpr} below.

\section{\label{secvalid} Conditions for the asymptotic expansions}

By considering the growing modes in the interior expansion for $\phi$, and
the singular modes in the exterior expansion, we can say something about the
position of the matching region, and in some cases show that a matching
region could not exist.

\subsection{Case I: McVittie background}

At $\mathcal{O}(\delta ^{2})$, the interior expansion of $\phi $ in the
McVittie metric has a mode that grows like $\xi $. This asymptotic expansion
will certainly cease to be valid if, when rewritten in some intermediate
scaling region, where $\xi \propto \delta ^{-\alpha }\xi $ say with $%
0<\alpha <1$, terms appear that scale as inverse powers of $\delta $. In
such a region $\delta ^{2}\xi \sim \delta ^{2-\alpha }\xi $, and so such
terms will always be suppressed by at least a single power of $\delta $. The
zeroth-order term, $\phi _{I}^{(0)}$, in the interior will itself have
ceased to be a valid asymptotic approximation to $\phi $ when the $\delta
^{2}\xi $ terms dominate over the $1/\xi $ term. This will occur only when $%
\xi \propto \delta ^{-1}$, i.e. in the exterior region. At order $\delta
^{4} $, the fastest growing mode will go like $\xi ^{3}$, and this will also
only conflict with the first two terms in the expansion when $\xi \propto
\delta ^{-1}$. We conclude that eq. (\ref{mcintexp}) is a valid asymptotic
approximation to $\phi $ everywhere outside the exterior region. Similarly
we note that the most singular term in the first two terms of the exterior
expansion, as given in Appendix \ref{appA}, goes like $\delta /\rho $. At $%
\mathcal{O}\left( \delta ^{2}\right) $ we would find terms that behave as $%
\delta ^{2}/\rho ^{2}$. The exterior expansion will therefore only cease to
be valid when $\rho \sim \delta $, i.e. in the interior region. Therefore,
in the McVittie case, the position of the intermediate region is not
important; we can choose any intermediate scaling and the matching procedure
will be valid. As a final check on our method, we have explicitly performed
the matching for the McVittie background in Appendix \ref{appA}, and shown
it to be self-consistent.

\subsection{\label{secval} Cases II and III: Tolman-Bondi models}

We assume that as $\mu _{1}\left( \chi \right) \sim \chi ^{n}$ as $\chi
\rightarrow \infty $ for some $n>0$. At order $\delta ^{q}$, the growing
mode in the interior approximation will grow like $\delta ^{q}\chi ^{n}/\xi $
in the Gautreau case, or as $\delta ^{q}\eta ^{n}/\xi $ for the simultaneous
big-bang models. Assuming that in some intermediate region $\eta ,\chi ,\xi
\sim \delta ^{-\alpha }$ with $0<\alpha <1$ we can see that this growing
mode will dominate over the zeroth order $1/\xi $ when $\alpha =q/n$, and
the asymptotic approximate will fail altogether if $n>1$ and $\alpha
>q/(n-1) $. We did not explicitly find the exterior expansion of $\phi $,
since we were only really concerned with its behaviour in the interior,
however the first non-homogeneous mode should behave as $\delta
^{p}z_{1}(\rho )/\rho $ if $p\leq 1$ or $\delta /\rho $ if $p>1$. If $%
z_{1}(\rho )\sim \rho ^{m}$ as $\rho \rightarrow 0$, then the exterior
expansion will break down if
\begin{eqnarray}
&m<1-p&\alpha \leq 1-\frac{p}{1-m}, \\
&m\geq 1-p&\alpha =0.  \notag
\end{eqnarray}%
In addition to this consideration, we note that the transformation $%
T\rightarrow T^{\ast }$ used in case III will not be well-defined if $\xi
\propto \delta ^{-\alpha }$ where $\alpha \geq 2/3$. Just by considering the
behaviour of the next-to-leading order terms we can say that if both the
interior and exterior zeroth-order approximations are to be simultaneously
valid in some intermediate scaling region we need, for the Gautreau case:
\begin{equation*}
\max \left( 0,1-\frac{p}{1-m}\right) <\alpha <\frac{q}{n},
\end{equation*}%
and for the simultaneous big-bang case:
\begin{equation*}
\max \left( 0,1-\frac{p}{1-m}\right) <\alpha <\min \left( \frac{q}{n}%
,2/3\right) .
\end{equation*}%
If such an intermediate region does indeed exist then the zeroth-order
matching performed in section \ref{zeroth} will be valid. The general form
of the interior approximation to $\mathcal{O}\left( \delta ^{p}\right) $
will then be correct; the only unknown function in this term is $B(\delta T)$%
. If the matching works to order $\delta ^{p}$, as well as zeroth-order,
then we have argued that $B(\delta T)$ will be quasi-static. If the matching
procedure does not work to this order then its quasi-static character may be
lost. However, we would not expect it to vary in time any faster than the
other $\mathcal{O}\left( \delta ^{p}\right) $ terms. So long as we can match
the zeroth-order approximations in some region, we can find the
circumstances under which condition (\ref{wettcond}) holds by comparing the
sizes of the two terms in (\ref{gatphitev}) and (\ref{ltbphitev}), for the
Gautreau and simultaneous big bang cases respectively.

\section{\label{intpr} Interpretation and Generalisation}

We assume that $2mF(\bar{\phi}_{0})/R_{s}\ll 1$ in the Tolman-Bondi cases II
and III considered above and that the matched asymptotic expansion method is
valid. Condition (\ref{wettcond}) will then only fail to hold if the $%
\mathcal{O}\left( \delta ^{q}\right) $ term in expressions (\ref{gatphitev})
and (\ref{ltbphitev}) is larger that the order $\delta \phi _{e}^{(0)\prime
} $ term. For condition (\ref{wettcond}) to hold we require:
\begin{equation}
\frac{B_{,\phi }(\phi _{c})\left( \int^{r}\mathrm{d}r^{\prime }\left( \frac{%
2m}{R(r^{\prime },t)}\right) ^{3/2}\frac{\delta ^{q}\mu _{1,r^{\prime
}}(r^{\prime })}{R(r,t)}+\left( 1-\frac{F(\bar{\phi}_{0})}{B(\phi _{c})}%
\right) \left( \frac{\left( 2m\right) ^{3/2}}{R}\right) \delta ^{q}\left(
r^{-1/2}\mu _{1,r}\right) _{R=R_{s}}\right) }{\dot{\phi}_{c}(t)}\ll 1,
\label{criteria}
\end{equation}%
where $R_{s}\chi ,R_{s}\eta \sim r$ and $R_{s}\xi =R$; $t-t_{0}=L_{I}T$. For
any given model this can be evaluated. As the expression is currently
written its physical meaning is somewhat obscured. Firstly, the lower limit
on the integral is not specified; we merely specified that the `constant' of
integration that arises from this lower limit should be quasi-static. We
should therefore take the lower limit to be some value of $r$ such that the
contribution to the integral from that lower limit vanishes. If $p>1$, the
above criterion (\ref{criteria}) will certainly hold. Therefore, in looking
for where (\ref{wettcond}) fails to hold, we are concerned only with the
cases where $p\leq 1$. In these situations, and indeed also for if $1<\delta
<2$, the local matter density close to $R=R_{s}$ will be much larger than
the cosmological dust density. We note that in the interior the energy density, $\epsilon (r,t)$  and also energy overdensity, $\Delta \epsilon $
(as $p<2$), is given by
\begin{equation*}
\kappa \epsilon (r,t)\sim \kappa \Delta \epsilon =\kappa (\epsilon -\epsilon
_{c})\sim 2m\delta ^{p}\frac{\mu _{1,r}}{R^{3/2}r^{1/2}}.
\end{equation*}%
We also note that $R _{,t}=\sim \pm (2m/R)^{1/2}$ and $%
R_{,r}\sim r^{1/2}/\xi ^{1/2}$. Therefore, in the interior region, we can substitute $\pm (2m/R)^{1/2}\mu _{1,r}$ for $%
\Delta(R_{,t} \kappa \epsilon) := \kappa(R_{,t}  \epsilon - HR \epsilon_c)$.  Doing this we find that our criterion, eq. (%
\ref{criteria}), becomes:
\begin{equation}
\left\vert \frac{B_{,\phi }(\phi _{c})\left( \int^{r}\mathrm{d}r^{\prime
}R_{,r} \Delta (\kappa R_{,t} \epsilon (r^{\prime },t))+\left( 1-\frac{F(\bar{%
\phi}_{0})}{B_{,\phi }(\phi _{c})}\right) \left( \xi _{,t}\kappa \Delta
\epsilon \right) _{R=R_{s}}\frac{R_{s}^{2}}{R}\right) }{\dot{\phi}_{c}(t)}%
\right\vert \ll 1.  \label{criteria2}
\end{equation}%
The two expressions, (\ref{criteria}) and (\ref{criteria2}), are equivalent
whenever the interior approximation holds. These should be checked in the
context of a given model. We can further simplify expression (\ref{criteria2}%
) by noting that, to zeroth order in $2m/R_{s}$, we should expect $F(\bar{%
\phi}_{0})\sim B_{,\phi }(\phi _{c})$, and so the second term in the
denominator will be a factor of $2m/\xi {s}$ smaller than the first and so
can safely be neglected.

Expression (\ref{criteria2}) is especially useful since it is written in
terms of the matter overdensity, $\kappa \epsilon $. Using this expression
we will now conjecture a more general criterion, that will coincide with (%
\ref{criteria}) and (\ref{criteria2}) whenever the matching procedure works,
but which we believe should also apply to many cases where the matching
would not have worked. Eq. (\ref{criteria2}) only features terms that are
well defined for \emph{all} Tolman-Bondi models, i.e. $R_{,r}$, $\xi ,{t}$
and $\kappa \Delta \epsilon $. The condition

\begin{equation}
\int_{\infty }^{r}\mathrm{d}r^{\prime }R_{,r}\Delta (R_{,t} \kappa \epsilon
(r^{\prime },t))<\infty  \label{infinteg}
\end{equation}%
will hold provided $\kappa \Delta \epsilon _{,t}\sim o\left( 1/R^{2}\right) $
as $R\rightarrow \infty $; this is certainly true whenever the extra mass
contained in the over-dense interior region is finite. The integrals in (%
\ref{criteria}) and (\ref{criteria2}) are specified by the condition that
any constant term that remains as $R\rightarrow \infty $ should be
neglected. So long as $\mu _{1,\chi }\propto \chi ^{n}$ with $n<5/2$, these integrals will be equivalent to the expression given in eq. (\ref%
{infinteg}), at least in the interior region. The analysis of section \ref%
{secval} suggests that if $n\geq 5/2$ and $q\leq 1$ the size of any
potential intermediate matching region will be strongly bounded; thus, for
most cases where we expect the matching procedure to work, $n<5/2$.

In the Tolman-Bondi models the radial velocity of the dust particles, as measured by an observer at fixed $R$, is given by $R_{,t}$. Assuming that the second term in denominator of (\ref{criteria})
and (\ref{criteria2}) is negligible compared to the first, we can rewrite
our criterion for condition (\ref{wettcond}) in the more concise and
revealing form by replacing $R_{,t}$ with the dust particle velocity, $v$:
\begin{equation}
\frac{\frac{2}{3}B_{,\phi }(\phi _{c})\int_{\gamma (R)}\mathrm{d}%
l\,\left\vert \Delta (v\kappa  \epsilon
)\right\vert }{\left\vert \dot{\phi}_{c}(t)\right\vert }\ll 1
\label{simpform}
\end{equation}%
where $\mathrm{d}l=\mathrm{d}r^{\prime }R_{,r}$ and the path of integration,
$\gamma (R)$, runs from $R$ to $\infty $, and $\Delta (v \kappa \epsilon)= v\kappa \epsilon - HR \epsilon_{c}$. Since we expect that $\phi $ will only feel the effects of events that
happened in its causal past, we should take $\gamma (R)$ to run from $R$ to
spatial infinity along a past, radially directed light-ray. For the
Tolman-Bondi models we have considered, the LHS of (\ref{simpform}) will
coincide (at least in magnitude) with the LHSs of (\ref{criteria}) and (%
\ref{criteria2}).

Eq. (\ref{simpform}) has the form that we might expect for condition (\ref%
{wettcond}) to hold inside a local, spherically-symmetric, inhomogeneous
region produced by an embedded Schwarzschild mass in an asymptotically FRW
universe: the LHS of eq. (\ref{simpform}) vanishes as we approach the
cosmological region and the local time variation of the dilaton field is
driven by the time variation of the energy density along its past light-cone.

\emph{We shall therefore conjecture that eq. (\ref{simpform}) is a general
sufficient condition for eq. (\ref{wettcond}) to hold: applicable even to
spherically-symmetric (or approximately spherically-symmetric), dust plus
cosmological constant, backgrounds in which the matching procedure would
itself fail. }

If the cosmic evolution of the dilaton field is dominated by its coupling to
matter, so $\left\vert B_{,\phi }(\phi _{c})\kappa \epsilon _{c}\right\vert
\gg \left\vert V_{,\phi}(\phi _{c})\right\vert $, then $\dot{\phi}_{c}(t)=%
\mathcal{O}\left( B_{,\phi }(\phi _{c})\kappa \epsilon _{c}(t)/H(t)\right) $
and condition (\ref{wettcond}) therefore holds, at the epoch $t=t_{0}$,
whenever
\begin{equation}
\int_{\gamma (R)}\mathrm{d}l\,H(t_{0})\frac{\left\vert \Delta (v \epsilon )\right\vert }{\epsilon _{c}(t_{0})}\ll 1.  \label{simpform2}
\end{equation}%
If, alternatively, the cosmic evolution is potential-driven, so $\left\vert
V_{,\phi }(\phi _{c})\right\vert \gg \left\vert B_{,\phi }(\phi _{c})\kappa
\epsilon _{c}\right\vert $, then the LHS of the above expression will be
suppressed by an additional factor of $\left\vert B_{,\phi }(\phi
_{c})\kappa \epsilon _{c}/V_{,\phi }(\phi _{c})\right\vert \ll 1$.

In section \ref{wettsec} we reviewed Wetterich's conclusion of \cite%
{wetterich:2002} that condition (\ref{wettcond}) holds true when the cosmic
evolution of the dilaton is potential dominated. Whilst we have identified
gaps in the analysis performed in that paper, we have been able to establish
a similar result. For a given evolution of the background matter density, we
have seen that condition (\ref{wettcond}) is more likely to hold (or will
hold more strongly) when $\left\vert B_{,\phi }(\phi _{c})\kappa \epsilon
_{c}/V_{,\phi }(\phi _{c})\right\vert \ll 1$. The domination by the
potential term in the cosmic evolution of the dilaton has a homogenising
effect on the time variation of $\phi $. In the final section we shall apply
eq. (\ref{simpform2}) to some physically relevant scenarios and show that,
on the basis of the above analysis, we should expect $\dot{\phi}(\mathbf{x}%
,t)\approx \dot{\phi}_{c}(t)$ in the solar system.

\section{Observations in our Solar System}

We would like to apply out results to solve our original problem: whether or
not condition (\ref{wettcond}) holds when the local region is the Earth or
the solar system. We will consider a star (and associated planetary system)
inside a galaxy that is itself embedded in a large galactic cluster. The
cluster is assumed to have virialised and be of size $R_{clust}$. Close to
the edge of the cluster we allow for some dust to be unvirialised and still
undergoing collapse. Since we have only performed our calculations for
spherically-symmetric backgrounds, we should, strictly speaking, also
require spherical symmetry about our star; for a realistic model this seems
contrived. In an upcoming paper, \cite{shawbarrow2}, we shall consider more fully the effect
that deviations from pure spherical symmetry have on our results; for now we
assume that these effects are small, and we can relax the spherical symmetry
requirement somewhat without invalidating our analysis. We consider the
different contributions to the LHS of condition (\ref{simpform2}) in this
astronomical set-up:
\begin{equation}
I:=\int_{\gamma (R)}\mathrm{d}l\,H(t_{0}) \, \frac{\left\vert \Delta (%
v \epsilon )\right\vert }{\epsilon _{c}(t_{0})}%
=I_{clust}+I_{gal}+I_{star}.  \label{I}
\end{equation}

\ \ For illustration we evaluate $I$ for Brans-Dicke theory. For other
theories, $I$ will still take very similar values. We assume that the
density of non-virialised matter, just outside the virialised cluster, is no
greater than the average density of the cluster and if we move away from the
edge of the virialised region, the over-density drops off quickly, i.e. as $%
R^{-s}$, $s>1/2$. In this case, the Tolman-Bondi result, eq. (\ref{ltbphitev}%
), applies and the magnitude of $I$ is bounded by:
\begin{eqnarray*}
I_{clust} &\lesssim &t_{0}^{-1}(s-1/2)^{-1}\sqrt{2M_{clust}R_{clust}}\frac{%
\epsilon _{clust}}{\epsilon _{c}} \\ \notag &\approx& \frac{1.2\times 10^{-7}}{\Omega_{m}(s-1/2)}\left(\frac{R_{clust}}{\mathrm{Mpc}}\right)\left(\frac{v_{clust}}{\mathrm{km}\,\mathrm{s}^{-1}}\right)(1+z_{vir})^{3} \\
&\approx &1.5 \times 10^{-3}[3/(2s-1)]\Omega _{m}^{-1}(1+z_{vir})^{3}\ll 1,
\end{eqnarray*}%
\noindent where we have used $3GM/R_{clust}=v_{clust}^{2}$. $R_{clust}$ is
the scale of the cluster post-virialisation, and $v_{clust}$ is the average
velocity of the virialised dust particles in it; $t_{0}=13.7Gyrs$ is the age of the universe. In the final approximation
we used representative values $R_{clust}=100Mpc$ and $v_{clust}=200\mathrm{%
km\,s}^{-1}$ appropriate for a cluster like Coma.  Taking a cosmological
density parameter equal to $\Omega _{m}=0.27,$ in accordance with WMAP, and $%
h=0.71$, we expect that for a typical cluster which virialised
at a redshift $z_{vir}\ll 1$, we would have $I_{clust}\approx 5.7\times
10^{-3}$. The term in $[]$ is unity when $s=2$, i.e. $2GM/r \rightarrow const$; such a matter distribution is characteristic of dark matter halos. Different
choices of $s>1/2$ can be seen to only change this estimate by an $\mathcal{O%
}(1)$ factor. As $s\rightarrow 1/2$ the matched asymptotic expansion method,
and hence this particular evaluation, breaks down. We believe that the
generalised formula for $I$ will, however, still give accurate results.
We have assumed that the baryon-to-dark matter ratio inside the cluster is
the same as in the universe on average. If this is not the case, and the
dilaton couples with vastly different strengths to baryons and dark matter,
then $I_{clust}$ could be made as large as $3 \times 10^{-2}$, for these values of $%
R_{clust}$ and $v_{clust}$; this scenario would require the cluster to be
comprised completely of baryons, and the dilaton to have zero coupling to
dark matter.

We now consider the galactic contribution. This will come about as a result
of the galaxy slowly accreting matter from the intergalactic medium (IGM).
We shall assume that the properties of IGM are the same as the average
properties of the cluster. In this case the density of the IGM will be
approximately constant and as such eq. (\ref{ltbphitev}) will not be
strictly applicable. However, if we assume that our conjecture of the
previous section holds then we can evaluate $I_{gal}$. The IGM has an
average particle velocity $v_{IGM}$; as a result of this, the gravitational
influence of our Galaxy upon it will only be felt significantly out to a
radius $R_{G}^{(gal)}=2GM_{gal}/v_{IGM}^{2}$. We therefore take $%
R=R_{G}^{(gal)}$, rather than $\infty $, as the upper limit of the integral
in eq. (\ref{simpform2}).
\begin{eqnarray*}
I_{gal} &\approx &-2t_{0}^{-1}\sqrt{2GM_{gal}R_{G}^{(gal)}}\frac{%
\epsilon _{IGM}}{\epsilon _{c}} \\
&= &-2.4\times 10^{-15}\Omega _{m}^{-1}\frac{M_{gal}}{M_{\odot }}%
(1+z_{vir})^{3}\left( \frac{\mathrm{km\,s}^{-1}}{v_{clust}}\right) \\ \notag &\approx
&-4.4 \times 10^{-6}(1+z_{vir})^{3}\ll 1, 
\end{eqnarray*}
\noindent where we have taken $\epsilon _{IGM}\approx \epsilon _{clust}$ and
$v_{IGM}\approx v_{clust}$. In the last line we have taken $%
M_{gal}=10^{12}M_{\odot }$ and $v_{clust}=200\mathrm{km\,s}^{-1}$ as above.

Finally we consider the contribution that results from our star accreting
matter from the interstellar medium (ISM). This calculation proceeds in the
same way as the one for $I_{gal}$.
\begin{eqnarray*}
I_{star} &\approx &-2t_{0}^{-1}\sqrt{2Gmr_{G}}\frac{\epsilon _{ISM}%
}{\epsilon _{c}} \\ \notag &=&-1.2\times 10^{-13}\Omega _{m}^{-1}\frac{m}{M_{\odot }}%
\left( \frac{\mathrm{km\,s}^{-1}}{v_{ISM}}\right) h^{-2}n \\
&\approx &-1.8\times 10^{-14}n\ll 1,
\end{eqnarray*}%
\noindent where $r_{G}=2Gm/v_{ISM}^{2}$, and $\epsilon _{ISM}=n$ $\mathrm{%
protons\,cm^{-3}}$ where $n\approx 1-10^{4}$. We have taken $v_{ISM}=5%
\mathrm{km\,s}^{-1}$ and $m=M_{\odot }$ to give the final approximation. It
is clear that, in general, $I_{star}\ll I_{gal}\ll I_{clust}$. The infall of
dust into cluster will tend to be the dominant contribution to the LHS of
eq. (\ref{simpform2}). The estimate of $I_{clust}$ given above should be
viewed as an upper bound on its value; even still we have seen that is small
compared with $1$, and hence that we should expect condition (\ref{wettcond}) to hold
near our star; with deviations from that behaviour bounded by the 0.6\% level (if the dilaton
couples only to baryonic matter this could rise to as high as 3\%). Assuming that the conditions in our solar system are not too
different from those considered above, we conclude that irrespective of the
value of the dilaton-to-matter coupling, and what dominates the cosmic
dilaton evolution, that
\begin{equation*}
\dot{\phi}(\mathbf{x},t)\approx \dot{\phi}_{c}(t)
\end{equation*}%
will hold in the solar system in general, and on Earth in particular. We
have seen that if the cosmic evolution of $\phi $ is potential dominated
then this only serves to strengthen this result. The only caveat is that
since we have ignored the back reaction of $\phi $ on the background
cosmological geometry we are limited to $\left\vert B_{,\phi }\right\vert
\ll 1$. If $B_{,\phi }$, and hence the back reaction, is large then the
scope for violations of condition (\ref{wettcond}) increases; if the
coupling to gravity of the dilaton is large, then dilaton field may itself
undergo gravitational collapse.

We can also conclude, from the above analysis, that, even if condition (\ref%
{wettcond}) is violated during the collapse of an overdense region of
matter, once the region stops collapsing as a result of its virialisation,
the time evolution of the dilaton field tends towards homogeneity. Contrary
to what has been claimed before in the literature, virialisation does \emph{%
not} stabilise the value of the dilaton, and protect it from any
cosmological variation.

In addition whilst we did not explicitly calculate the interior solution for
$2m/R_{s}=\mathcal{O}(1)$, the evolution eq. (\ref{phiintev}) is still valid
in this case. We should therefore expect the magnitude of the rate of time
variation in $\phi _{I}^{(1)}$ to be similar to what was found for the $%
2m/R_{s}\ll 1$ case. We therefore conclude that over time-scales that are
large compared to $2Gm$, condition (\ref{wettcond}) will hold whenever
condition (\ref{simpform2}) does, and there will be \emph{no} significant
gravitational memory of the value of $G$ from the epoch when black holes or
gravitationally bound structures first formed. This result agrees with the
prediction made by Jacobson, \cite{jacobson}, the inhomogeneous models of
\cite{mem2}, and the numerical calculations of Harada et. al., \cite{harada}
, who also studied the Tolman-Bondi background. Indeed we have seen that
whenever the, not very restrictive, condition, $\epsilon _{l}/\epsilon
_{c}\ll 1/(2mH_{0})(\gg 1)$, holds then eq. (\ref{phijac}), gives a true
asymptotic approximation to the behaviour of the dilaton close to the
horizon.

\section{Summary}

In this paper we have considered the extent to which a cosmological time
variation in a scalar field (dilaton) would be detectable on the surface of
gravitational-bound systems that are otherwise disconnected from changes
that occur over cosmological scales. This problem is of particular relevance
when the dilaton defines the local value of one of the traditional
`constants' of Nature, and when the system is the Earth or our solar system.
Several scalar-tensor theories have already been developed which
self-consistently describe the variations of traditional `constants' of
Nature, like $\alpha $ and $G$. By matching rigorously constructed
asymptotic expansions of the associated scalar field found in different
limits, close to the Earth and far away from it at cosmological scales, we
have been able to derive approximate, analytical expressions for the scalar
field near the surface of such bound systems. This result was found under
two major assumptions:  the physically realistic condition that the scalar
field should be weakly coupled to matter and gravity(in effect the
variations of `constants' on large scales occur more slowly than the
universe is expanding )  and thus have a negligible back-reaction on the
cosmological background, and the less realistic one of spherical symmetry.
We do not expect the relaxation of the spherical symmetry condition to
greatly alter the qualitative nature of our conclusions, and we shall
present a rigorous treatment of the problem when no symmetry is present in a
subsequent work.

Finally we have extracted from our analysis a sufficient condition for the
local time-variation of a scalar field, or varying physical `constant', to
track the cosmological one, and we have proposed a generalisation of this
condition that is applicable to scenarios more general than those explicitly
considered here. By evaluating the condition for an astronomical scenario
similar to the one appropriate for our solar system, we have concluded that
almost irrespective of the form of the dilaton-to-matter, and the form of
the dilaton self-interaction, its time variation in the solar system will
track the cosmological one. We have therefore provided a general proof of
what was previously merely assumed: that \emph{terrestrial} and \emph{solar
system} based observations can legitimately, be used to constrain the \emph{%
cosmological} time variation of supposed `constants' of Nature. \newline

\noindent \textbf{Acknowledgements} DS is supported by a PPARC studentship.
We would like to thank P.D. D'Eath and T. Clifton for helpful discussions. \appendix

\section{\label{appA} Explicit matching for the McVittie background}

\subsection{Exterior Solution}

In the exterior region $\phi _{E}\sim \phi _{E}^{(0)}+\delta \phi _{E}^{(1)}+%
\mathcal{O}(\delta ^{2})$ where:
\begin{eqnarray}
\frac{1}{\tilde{a}^{3}}\partial _{\tau }\left( \tilde{a}^{3}\partial _{\tau
}\phi _{E}^{(1)}\right)  &-&\frac{\left( 1+\tfrac{1}{4}\Omega _{k}^{0}\rho
^{2}\right) ^{3}}{\tilde{a}^{2}\rho ^{2}}\partial _{\rho }\left( \frac{\rho
^{2}}{1+\tfrac{1}{4}\Omega _{k}^{0}\rho ^{2}}\partial _{\rho }\phi
_{E}^{(1)}\right)  \\
&=&\left( \frac{2m}{R_{s}}\right) ^{3/2}\frac{(1+\tfrac{1}{4}\Omega _{k}\rho
^{2})^{1/2}}{\tilde{a}(\tau )\rho }\left\{ \phi _{E}^{(0)\prime \prime
}+h\phi _{E}^{(0)\prime }+3B_{,\phi }(\phi _{E})\left( h^{\prime }+\tfrac{11%
}{4}\Omega _{k}(\tau )\right) \right\}   \notag \\
&&+B_{,\phi \phi }(\phi _{E}^{(0)})\kappa \epsilon _{dust}^{0}\phi
_{E}^{(1)}-V_{,\phi \phi }(\phi _{E}^{(0)})\phi _{E}^{(1)}.  \notag
\end{eqnarray}%
Despite the complexity of this formula we can solve it admits separable
solutions in $\tau $ and $\rho $:
\begin{equation*}
\phi _{E}^{(1)}=\left( \frac{2m}{R_{s}}\right) ^{3/2}\frac{\left( 1+\frac{1}{%
4}\Omega _{k}^{0}\rho ^{2}\right) ^{1/2}}{\rho }\Upsilon (\tau )+\Phi (\tau
\rho ).
\end{equation*}%
\noindent where $\Upsilon (\tau )$ satisfies the ODE:
\begin{eqnarray*}
\frac{1}{\tilde{a}^{3}}\partial _{\tau }\left( \tilde{a}^{3}\partial _{\tau
}\Upsilon \right)  &-&\frac{3\Omega _{k}(\tau )}{4}\Upsilon -B_{,\phi \phi
}(\phi _{E}^{(0)})H_{0}^{-2}\kappa \epsilon _{dust}^{0}\Upsilon
+H_{0}^{-2}V_{,\phi \phi }(\phi _{E}^{(0)})\Upsilon  \\
&=&-\frac{1}{\tilde{a}}\left\{ \phi _{E}^{(0)\prime }+h\phi _{E}^{(0)\prime
}+3B_{,\phi }(\phi _{E})\left( h^{\prime }+\tfrac{11}{4}\Omega _{k}(\tau
)\right) \right\} .
\end{eqnarray*}%
When $k>0$, $\Phi =\sum_{\gamma =-\infty }^{\infty }\sqrt{\Omega _{k}^{0}}%
\mathbf{c}_{\gamma }\cdot \mathbf{T}_{\gamma }^{+}(\tau )X_{\gamma
}^{+}(\rho )$, $\mathbf{c}_{\gamma }\in \mathbb{C}^{2}$:
\begin{eqnarray*}
X_{\gamma } &=&\frac{e^{i\gamma \alpha (\rho )}}{\sin \alpha } \\
\alpha (\rho ) &=&\sin ^{-1}\left( \frac{\sqrt{\Omega _{k}^{0}}\rho }{1+%
\tfrac{1}{4}\Omega _{k}^{0}\rho ^{2}}\right) ,
\end{eqnarray*}%
and $\mathbf{T}_{\gamma }^{+}=\left( T_{\gamma }^{+,1},T_{\gamma
}^{+,2}\right) ^{t}$, where $T_{\gamma }^{+,1}$ and $T_{\gamma }^{+,2}$ are
linearly independent solutions of the following ODE:
\begin{equation*}
\frac{1}{\tilde{a}^{3}}\partial _{\tau }\left( \tilde{a}^{3}\partial _{\tau
}T_{\gamma }^{+,i}(\tau )\right) +(\gamma ^{2}-1)\Omega _{k}(\tau )T_{\gamma
}^{+,i}(\tau )=B_{,\phi \phi }(\phi _{E}^{(0)})H_{0}^{-2}\kappa \epsilon
_{dust}^{0}T_{\gamma }^{+,i}(\tau )-H_{0}^{-2}V_{,\phi \phi }(\phi
_{E}^{(0)})T_{\gamma }^{+,i}(\tau ).
\end{equation*}%
If $k=0$. then $\Phi =\int_{-\infty }^{\infty }\mathrm{d}\gamma \,\mathbf{c}%
(\gamma )\cdot \mathbf{T}_{\gamma }^{0}(\tau )X_{\gamma }^{0}(\rho )$, $%
\mathbf{c}(\gamma )\in \mathbb{C}^{2}$, and $X_{\gamma }^{0}(\rho
)=e^{i\gamma \rho }/\gamma \rho $. $\mathbf{T}_{\gamma }^{0}(\tau )=\left(
T_{\gamma }^{0,1},T_{\gamma }^{0,2}\right) ^{t}$ satisfies:
\begin{equation*}
\frac{1}{\tilde{a}^{3}}\partial _{\tau }\left( \tilde{a}^{3}\partial _{\tau
}T_{\gamma }^{0,i}(\tau )\right) +\gamma ^{2}T_{\gamma }^{0,i}(\tau
)=B_{,\phi \phi }(\phi _{E}^{(0)})H_{0}^{-2}\kappa \epsilon
_{dust}^{0}T_{\gamma }^{0,i}(\tau )-H_{0}^{-2}V_{,\phi \phi }(\phi
_{E}^{(0)})T_{\gamma }^{0,i}(\tau ),
\end{equation*}%
\noindent where $i=1,2$. Finally if $k<0$ then $\Phi =\int_{-\infty
}^{\infty }\mathrm{d}\gamma \,\sqrt{-\Omega _{k}^{0}}\mathbf{c}(\gamma
)\cdot \mathbf{T}_{\gamma }^{-}(\tau )X_{\gamma }^{-}(\rho )$, $\mathbf{c}%
(\gamma )\in \mathbb{C}^{2}$:
\begin{eqnarray*}
X_{\gamma }^{-}(\rho ) &=&\frac{e^{i\gamma \alpha }}{\sinh \alpha } \\
\alpha  &=&\sinh ^{-1}\left( \frac{\sqrt{-\Omega _{k}^{0}}\rho }{1+\tfrac{1}{%
4}\Omega _{k}^{0}\rho ^{2}}\right) ,
\end{eqnarray*}%
and $\mathbf{T}_{\gamma }^{-}(\tau )=\left( T_{\gamma }^{-,1},T_{\gamma
}^{-,2}\right) ^{t}$:
\begin{equation*}
\frac{1}{\tilde{a}^{3}}\partial _{\tau }\left( \tilde{a}^{3}\partial _{\tau
}T_{\gamma }^{-,i}(\tau )\right) -\left( \gamma ^{2}+1\right) \Omega
_{k}(\tau )T_{\gamma }^{-,i}(\tau )=B_{,\phi \phi }(\phi
_{E}^{(0)})H_{0}^{-2}\kappa \epsilon _{dust}^{0}T_{\gamma }^{-,i}(\tau
)-H_{0}^{-2}V_{,\phi \phi }(\phi _{E}^{(0)})T_{\gamma }^{-,i}(\tau ),
\end{equation*}%
\noindent where $i=1,2$. In all cases we shall fix our definition of the $%
T_{\gamma }$ by the normalisation: $T_{\gamma }(\tau _{0})=1$.

\subsection{Matching Conditions}

By making our interior and exterior solutions for $\phi$ we see that: $%
\mathbf{c}_{\gamma} = \bar{\mathbf{c}}_{-\gamma}$ for $k>0$, and $\mathbf{c}%
(\gamma) = \bar{\mathbf{c}}(-\gamma)$; $\bar{z}$ is the complex conjugate of
$z$. Defining $A_{n}(\tau) = \sum_{\gamma} \gamma^{n} \mathbf{c}_{\gamma}
\cdot \mathbf{T}_{\gamma}^{+}$ or $A_{n}(\tau) = \int \mathrm{d}{\gamma}\;
\gamma^{n} \mathbf{c}({\gamma}) \cdot \mathbf{T}_{\gamma}^{0 / -}$ for $k > 0
$ and $k \leq 0$ respectively we see that:
\begin{eqnarray}
A_{0}(\tau) &=& \frac{F(\bar{\phi}_{0})}{a} - \Upsilon(\tau) \\
A_{1}(\tau) &=& 0 \Leftrightarrow c(\gamma) = \bar{c}(-\gamma) \\
\left \vert \Omega_{k}^{0} \right \vert A_{2}(\tau) &=& (h^{\prime}+ 2h^2)%
\tilde{a} F(\bar{\phi}_{0}) + \left(B_{,\phi \phi}(\phi_{E}^{(0)})\kappa
\epsilon_{dust}^{0} - V_{,\phi \phi}\right) \tilde{a} F(\bar{\phi}_{0}) +
\Omega_{k}(\tau) \tilde{a} F(\bar{\phi}_{0}) \\
&&- \tilde{a}\left(\phi_{E}^{(0)\prime\prime} + h\phi_{E}^{(0)\prime} +
3B_{,\phi}\left(h^{\prime}+ \tfrac{11}{4}\Omega_{k}(\tau)\right)\right) -
\frac{1}{4}\Omega_{k}^{0}\Upsilon.  \notag
\end{eqnarray}
In principle we can invert the equation for $A_{0}(\tau)$ to find the $%
\mathbf{c})(\gamma)$.

Given the definition of the $A_{n}$, the expressions for $A_{0}$ and $A_{2}$
must satisfy a consistency relation (if there were not to satisfy this, the
matching procedure would be invalid). We have checked that this relation
does \emph{indeed} hold here. The relation is:
\begin{equation*}
\left\vert \Omega _{k}(\tau )\right\vert A_{2}-\Omega _{k}(\tau )A_{0}=-%
\frac{1}{\tilde{a}^{3}}\partial _{\tau }\left( \tilde{a}^{3}\partial _{\tau
}A_{0}\right) +\left( B_{,\phi \phi }(\phi _{E}^{(0)})\kappa \epsilon
_{dust}^{0}-V_{,\phi \phi }\right) A_{0}.
\end{equation*}%
We can see, explicitly, that the matched asymptotic expansion method works
for the McVittie background.


\begin{thebibliography}{99}
\bibitem{webb} J.K. Webb et al, Phys. Rev. Lett. \textbf{82}, 884 (1999); M.
T. Murphy et al,\ Mon. Not. Roy. Astron. Soc. \textbf{327}, 1208 (2001);
J.K. Webb et al, Phys. Rev. Lett. \textbf{87}, 091301 (2001); M.T. Murphy,
J.K. Webb and V.V. Flambaum, Mon. Not R. astron. Soc. \textbf{345}, 609
(2003).

\bibitem{chand} H. Chand et al., Astron. Astrophys. \textbf{417}, 853
(2004); R. Srianand et al., Phys. Rev. Lett. \textbf{92}, 121302 (2004).

\bibitem{sdss} J. Bahcall, C.L. Steinhardt, and D. Schlegel, Astrophys. J.%
\textbf{\ 600}, 520 (2004).

\bibitem{qu} R. Quast, D. Reimers and S.A Levshakov, A. \& A. \textbf{386},
796 (2002).

\bibitem{lev} S.A. Levshakov, \textit{et al.,} A. \& A. \textbf{434}, 827
(2005).

\bibitem{lev2} S.A. Levshakov, \textit{et al.,} astro-ph/0511765.

\bibitem{rocha} G. Rocha, R. Trotta, C.J.A.P. Martins, A. Melchiorri, P.P.
Avelino, R. Bean, and P.T.P. Viana, Mon. Not. astron. Soc. \textbf{352,} 20
(2004).

\bibitem{darl} J. Darling, Phys. Rev. Lett. \textbf{91}, 011301 (2003).

\bibitem{oh} J. Darling, Astrophys. J. \textbf{612,} 58 (2004).

\bibitem{drink} M.J. Drinkwater, J.K. Webb, J.D. Barrow and V.V. Flambaum,
Mon. Not. R. Astron. Soc. \textbf{295}, 457 (1998).

\bibitem{ubachs} W. Ubachs and E. Reinhold, Phys. Rev. Lett. \textbf{92},
101302 (2004).

\bibitem{petit} R. Petitjean et al, Comptes Rendus Acad. Sci. (Paris)
\textbf{5}, 411 (2004).

\bibitem{tz} P. Tzanavaris, J.K. Webb, M.T. Murphy, V.V. Flambaum, and S.J.
Curran, Phys.Rev.Lett. \textbf{95}, 041301 (2005).

\bibitem{bert} B. Bertotti, L. Iess and P. Tortora, Nature \textbf{425}, 374
(2003).

\bibitem{uzan} J.P. Uzan, Rev. Mod. Phys. J.P. Uzan, Rev. Mod. Phys.\textbf{%
\ 75}, 403 (2003): J.-P. Uzan, astro-ph/0409424.

\bibitem{olive} K.A. Olive and Y-Z. Qian, Physics Today, pp. 40-5 (Oct.
2004).

\bibitem{jdb} J.D. Barrow, \textit{The Constants of Nature: from alpha to
omega}, (Vintage, London, 2002).

\bibitem{jdbroysoc} J.D. Barrow, Phil. Trans. Roy. Soc. \textbf{363}, 2139
(2005).

\bibitem{posp} K.A. Olive and M. Pospelov, Phys. Rev. D \textbf{65,} 085044
(2002); E. J. Copeland, N. J. Nunes and M. Pospelov, Phys.Rev. D \textbf{69,}
023501 (2004); S. Lee, K.A. Olive and M. Pospelov, Phys.Rev. D \textbf{70,}
083503 (2004); G. Dvali and M. Zaldarriaga, hep-ph/0108217; T. Damour and A.
Polyakov, Nucl. Phys. B \textbf{423}, 532 (1994); P.P. Avelino, C.J.A.P.
Martins, and J.C.R.E. Oliveira, Phys.Rev. D \textbf{70} 083506 (2004).

\bibitem{bd} C. Brans and R.H. Dicke, Phys. Rev. \textbf{124}, 925 (1961).

\bibitem{bkm} J.D. Barrow, D. Kimberly and J. Magueijo, Class. Quantum Grav.
\textbf{21}, 4289 (2004).

\bibitem{bek} J.D. Bekenstein, Phys. Rev. D \textbf{25,} 1527 (1982).

\bibitem{bsbm} H. Sandvik, J.D. Barrow\ and J. Magueijo, Phys. Rev. Lett.
\textbf{88}, 031302 (2002); J.~D.~Barrow, H.~B.~Sandvik and J.~Magueijo,
Phys.\ Rev.\ D \textbf{65}, 063504 (2002); J.~D.~Barrow, H.~B.~Sandvik and
J.~Magueijo, Phys.\ Rev.\ D \textbf{65}, 123501 (2002); J.~D.~Barrow,
J.~Magueijo and H.~B.~Sandvik, Phys. Rev. D \textbf{66}, 043515 (2002);
J.~Magueijo, J.~D.~Barrow and H.~B.~Sandvik, Phys. Lett. B \textbf{541}, 201
(2002); H. Sandvik, J.D. Barrow and J. Magueijo, Phys. Lett. B \textbf{549,}
284 (2002).

\bibitem{bmmu} J.D. Barrow and J. Magueijo, Phys. Rev. D \textbf{72}, 043521
(2005).

\bibitem{bs} D. Kimberly and J. Magueijo, Phys. Lett. B \textbf{584,} 8
(2004).

\bibitem{sb} D. Shaw and J.D. Barrow, Phys. Rev. D \textbf{71}, 063525
(2005).

\bibitem{newtgrav} J.D. Barrow, Mon. Not. R. astr. Soc., \textbf{282}, 1397
(1996).

\bibitem{bmot} D. Mota and J.D. Barrow, Mon. Not. Roy. Astron. Soc. \textbf{%
349}, 281 (2004); J.D. Barrow and D. Mota, Phys. Lett. B \textbf{581}, 141(
2004); T. Clifton, D. Mota and J.D. Barrow, Mon. Not. R. Astron. Soc.
\textbf{358}, 601 (2005).

\bibitem{jbspace1} J.D. Barrow and C. O'Toole, Mon. Not. Roy. Astron. Soc.
\textbf{322}, 585 (2001).

\bibitem{jbspace} J.D. Barrow, Phys. Rev. D \textbf{71}, 083520 (2005).

\bibitem{wetterich:2002} C. Wetterich, JCAP \textbf{10}, 002 (2003).

\bibitem{mem1} J.D. Barrow, Phys. Rev. D \textbf{46}, R3227 (1992); J.D.
Barrow, Gen. Rel. Gravitation \textbf{26}, 1 (1992); J.D. Barrow and B.J.
Carr, Phys. Rev. D \textbf{54}, 3920 (1996).

\bibitem{mem2} H. Saida and J. Soda, Class. Quantum Gravity \textbf{17},
4967 (2000); J.D. Barrow and N. Sakai, Class. Quantum Gravity \textbf{18},
4717 (2001).

\bibitem{harada} T. Harada, C. Goymer and B. J. Carr, Phys. Rev. D. \textbf{%
66}, 104023 (2002).

\bibitem{jacobson} T. Jacobson, Phys. Rev. Lett. \textbf{83}, 2699 (1999).

\bibitem{Cole} J. D. Cole, \emph{Perturbation methods in applied mathematics}%
, (Blaisdell, Waltham, Mass., 1968).

\bibitem{Hinch} E. J. Hinch, \emph{Perturbation methods}, (Cambridge UP,
Cambridge, 1991).

\bibitem{burkethorne} W. L. Burke and K. Thorne, in \emph{Relativity},
edited by M. Carmeli, S. Fickler, and L. Witten (Plenum Press, New York and
London, 1970), pp. 209-228.

\bibitem{burke} W. L. Burke, J. Math. Phys. \textbf{12}, 401 (1971).

\bibitem{Death1} P. D. D'Eath, Phys. Rev. D. \textbf{11}, 1387 (1975).

\bibitem{Death2} P. D. D'Eath, Phys. Rev. D. \textbf{12}, 2183 (1975).

\bibitem{Geroch} R. Geroch, Commun. Math. Phys. \textbf{13}, 180 (1969).

\bibitem{kr} A. Krasinski, \textit{Inhomogeneous Cosmological Models},
(Cambridge UP, Cambridge, 1996).

\bibitem{Gautreau:1984} R. Gautreau, Phys. Rev. D. \textbf{29}, 198 (1984).

\bibitem{mcv} G.C. McVittie, Mon. Not. Roy. Astron. Soc. \textbf{93}, 325
(1933); G.C. McVittie, \emph{General Relativity and Cosmology} (Chapman and
Hall, London, 1965).

\bibitem{Leibovitz} C. Leibovitz, Phys. Rev. D \textbf{4}, 2949 (1971).

\bibitem{barrowst} J. D. Barrow and J. Stein-Schabes, Phys. Lett. \textbf{%
103A}, 315 (1984).

\bibitem{barrowindia} U. Debnath, S. Chakraborty, and J.D. Barrow, Gen. Rel.
Gravitation \textbf{36}, 231 (2004).

\bibitem{shawbarrow2} D.  J. Shaw and J. D. Barrow, gr-qc/0601056.
\end{thebibliography}
\end{document}